\documentclass[journal]{IEEEtran}
\usepackage{amsmath,amsthm,graphicx,amsfonts,amssymb,epsfig,subfig,mathrsfs,mathtools,esvect,gensymb, xcolor}
\usepackage{xfrac}
\usepackage{xurl}
\usepackage{wrapfig}
\usepackage{arydshln}
\usepackage{blkarray}

\usepackage{color,soul}
\usepackage[normalem]{ulem}
\usepackage{cancel}

\usepackage[final]{changes} \definecolor{myblue}{RGB}{0,0,255} \definechangesauthor[name=Sungjun,color=myblue]{SJ}

\usepackage{multirow}
\usepackage{multicol}
\usepackage{lipsum}
\usepackage{dblfloatfix}
\usepackage{algorithm}
\usepackage{algpseudocode}
\usepackage{tikz}

{}
{}
\newtheorem{remark}{Remark}{}
\newtheorem{assumption}{Assumption}{}
{}
\newtheorem{proposition}{Proposition}{}
\newtheorem{theorem}{Theorem}{}

\ifCLASSINFOpdf
\else
\fi
\begin{document}
%
\title{Density-Driven Multi-Agent Coordination for Efficient Farm Coverage and Management in Smart Agriculture}
%
%
%

\author{Sungjun~Seo,~\IEEEmembership{Member,~IEEE,}
        and~Kooktae~Lee,~\IEEEmembership{Member,~IEEE}

\thanks{S. Seo and K. Lee are with the Department of Mechanical Engineering, 
New Mexico Institute of Mining and Technology, Socorro, NM 87801, USA 
(e-mail: sungjun.seo@student.nmt.edu; kooktae.lee@nmt.edu).}

\thanks{This is the author's accepted manuscript (AAM) of a paper accepted 
for publication in the IEEE Transactions on Control Systems Technology (TCST). 
The final version will be available on IEEE Xplore upon publication. 
© 2025 IEEE.}
}

\maketitle


\begin{abstract} 
The growing scale of modern farms has led to an increased demand for efficient and adaptive strategies for pest, weed, and disease management. Traditional methods, such as manual inspections and blanket pesticide spraying, often result in excessive chemical use, resource wastage, and environmental degradation. Unmanned aerial vehicles (UAVs) have emerged as a promising solution for precision agriculture, offering the potential for targeted spraying that enhances operational efficiency. However, existing UAV-based approaches face limitations, including restricted battery life, small payload capacity, and inefficient large-area coverage, particularly when relying on single-UAV systems. Multi-UAV coordination strategies have been proposed to address these challenges, yet many of these frameworks still rely on uniform spraying, neglecting important factors such as infestation severity, UAV dynamics, and energy efficiency.

This paper introduces a Density-Driven Optimal Control (D$^2$OC) framework, leveraging Optimal Transport (OT) theory to optimize UAV-based pesticide spraying in large-scale agricultural environments. The proposed method facilitates adaptive resource allocation based on infestation severity, reducing unnecessary chemical application. UAVs are modeled as a linear time-varying (LTV) system, accounting for variations in mass and moment of inertia due to payload changes. The D$^2$OC-based control strategy, derived through Lagrangian mechanics, ensures efficient fleet coordination, balanced workload distribution, and optimized mission duration. Simulation results demonstrate that the proposed approach outperforms traditional methods, such as uniform coverage and Spectral Multi-scale Coverage (SMC), in terms of coverage efficiency, chemical reduction, and sustainability, providing a scalable and resource-efficient solution for smart farming. 
\end{abstract}


%
\IEEEpeerreviewmaketitle

\section{Introduction}

The integration of advanced technologies into modern agriculture has revolutionized farming practices, significantly enhancing efficiency, sustainability, and precision. The advent of automation, data analytics, and connected systems has enabled optimized resource management, allowing farmers to make data-driven decisions that improve productivity and reduce waste. Real-time sensor networks now provide continuous monitoring of key environmental factors such as soil moisture, temperature, and crop health, facilitating adaptive interventions. Moreover, artificial intelligence (AI) has empowered predictive analytics for yield estimation, pest detection, and irrigation optimization, reducing dependency on manual labor while increasing overall efficiency \cite{khanal2024iot, goap2018iot}. 

However, despite these advancements, large-scale farms still face persistent challenges in pest, weed, and disease control. Conventional management approaches, such as manual inspection and uniform spraying, are not only labor-intensive but also inefficient, leading to excessive pesticide use, resource wastage, and environmental contamination. The increasing scale of modern farms, often spanning hundreds of acres, exacerbates these inefficiencies, necessitating adaptive and scalable solutions.

In this context, unmanned aerial vehicles (UAVs) have emerged as a promising tool for precision agriculture, particularly in targeted spraying applications. UAV-based spraying offers improved efficiency, precision, and reduced chemical waste compared to traditional methods. However, single-UAV systems are constrained by limited battery life and small payload capacity, which significantly restrict their operational time and coverage area. The need for regular battery recharging or replacement creates substantial downtime, especially in large-scale agricultural operations. Since UAVs typically rely on battery-powered energy sources, they are subject to battery depletion after a limited number of flight hours, often requiring several recharging cycles throughout the day. This limitation results in inefficient operation, especially when dealing with vast agricultural fields that require continuous spraying. Furthermore, small payload capacities restrict the amount of chemicals a single UAV can carry, forcing frequent landings for reloading, further contributing to operational downtime \cite{yu2019coverage}. To address these challenges, multi-UAV systems have been proposed to distribute spraying tasks among multiple drones, thus improving coverage efficiency, reducing downtime, and enhancing the overall sustainability of spraying operations. Despite these advantages, existing UAV-based solutions still rely on uniform blanket spraying, which can lead to excessive chemical use, soil degradation, and water contamination. More advanced strategies are needed to optimize UAV coordination, minimize chemical waste, and ensure effective pest and disease management.

\subsection{Literature Survey}

UAV deployment in smart agriculture has garnered significant attention, with research primarily focusing on optimizing spraying efficiency, monitoring, and coverage strategies. Rejeb et al. \cite{rejeb2022drones} conducted a bibliometric analysis of UAV applications in agriculture, emphasizing their roles in soil condition estimation, crop monitoring, and other precision agricultural tasks. Their study highlights the advantage of UAVs in providing higher spatial resolution than satellite systems, delivering more accurate and timely data for agricultural management. Despite these benefits, they also point to challenges such as the complexity of data processing, integration with existing agricultural practices, and limitations related to cost, regulation, and the need for effective training to maximize UAV capabilities.

Vision-based approaches, including convolutional neural networks (CNNs) for plant health assessment \cite{kamilaris2018deep} and object detection for crop stress identification \cite{li2022design}, further enhance monitoring capabilities. Yet, these methods are often limited by their failure to integrate with multi-UAV coordination, which constrains their scalability for large-area applications. This integration gap between vision-based monitoring and multi-UAV systems represents a critical research opportunity, as it is essential for efficient deployment across extensive agricultural fields.

To tackle such challenges, recent studies have proposed several UAV coordination strategies aimed at enhancing efficiency in large-scale agricultural tasks. Tevyashov et al. \cite{tevyashov2022algorithm} introduced a trajectory optimization algorithm that minimizes flight time based on predefined paths. While effective in controlled environments, this approach lacks adaptability to dynamic farmland conditions, such as terrain variations and environmental obstacles. As a result, UAVs may encounter inefficiencies when faced with unexpected changes, such as obstacles or unpredictable weather. Similarly, Liang and Delahaye \cite{liang2019drone} investigated UAV fleet deployment for large-scale surveying under the assumption of stable communication between UAVs. Unfortunately, in remote agricultural areas, communication reliability is often compromised, which could lead to system failure in real-world applications. This highlights the importance of decentralized communication protocols to ensure the robustness and reliability of multi-UAV systems in agricultural environments.

Further advancing UAV coordination, Skobelev et al. \cite{skobelev2018designing} developed a swarm-based system that facilitates decentralized decision-making among UAVs. While this approach supports greater flexibility in coordination, it does not address task allocation based on the priority of different areas in the field. In practical applications, certain regions may require more intensive spraying due to pest infestations, weed growth, or disease outbreaks, demanding priority-aware coordination among UAVs.

Despite the potential advantages of priority-aware spraying, most existing systems still rely on uniform blanket spraying or non-differentiated coverage, which can result in excessive chemical usage in areas that do not require treatment. This not only leads to inefficient resource use but also contributes to unnecessary environmental harm. The ability to prioritize spraying based on real-time field conditions, such as pest severity, weed density, or disease spread, is crucial for improving resource efficiency and minimizing environmental impact. Din et al. \cite{din2022deep} explored the use of deep reinforcement learning (DRL) for multi-UAV coverage, proposing a system that adapts to dynamic conditions. While this method incorporates environmental feedback, it does not fully address priority-based resource allocation, which is essential for optimizing spraying efficiency. Similarly, Ming et al. \cite{ming2023comparative} analyzed various UAV swarm control strategies aimed at improving coverage efficiency and flight path coordination. Although their work provides valuable insights into swarm control, it does not consider mechanisms to prioritize spraying based on the urgency of treatment in different field areas.

The integration of priority-aware spraying into multi-UAV systems requires sophisticated coordination mechanisms. Roque-Claros et al. \cite{roque2024uav} proposed a machine-learning-based swarm intelligence model for path planning and navigation optimization. While this model improves UAV coordination for effective area coverage, it lacks the capacity to prioritize regions with critical needs. Furthermore, its focus on path optimization overlooks the varying levels of treatment required across the field. Integrating priority-aware spraying within the UAV coordination framework would enable better resource targeting, ensuring that UAVs focus their time and spraying capabilities on areas with the most pressing needs.

Energy efficiency is another crucial consideration for large-scale UAV operations, as it allows sustained coverage without frequent recharging or battery replacement. While many strategies, such as energy-efficient path planning or battery swapping, contribute to prolonged mission duration, they often take a backseat to the more immediate need for priority-aware spraying in many agricultural scenarios. While energy management is important, it should not overshadow task prioritization, which is key to enhancing the effectiveness of UAV systems. Skobelev et al. \cite{skobelev2018designing} acknowledge the importance of energy consumption in their swarm-based system for agricultural surveying. However, their work is primarily focused on improving area coverage rather than task prioritization. Thus, to fully optimize the efficiency of multi-UAV systems in agriculture, it is essential to integrate energy optimization with priority-aware spraying, ensuring that UAVs target critical areas while managing energy usage effectively.

While existing UAV coordination and path planning strategies have made notable advancements in improving coverage and operational efficiency, a significant gap remains in the incorporation of priority-aware spraying. The ability to allocate UAV resources based on the severity of pests, weed density, or disease spread is crucial for optimizing spraying operations. Future systems must also integrate energy efficiency with priority-based task allocation to ensure both sustainable and targeted interventions. Currently, most existing frameworks lack a holistic approach that balances spraying efficiency, energy consumption, and sustainability, which are key factors necessary for large-scale deployments in smart agriculture.

\subsection{Contributions}

This paper introduces a novel multi-drone coordination strategy for large-scale agricultural pest, weed, and disease management. The proposed Density-Driven Optimal Control (D$^2$OC) framework, rooted in Optimal Transport (OT) theory, enables UAVs to allocate spraying resources adaptively based on the severity of infestation while minimizing unnecessary chemical application. Unlike conventional methods, this approach explicitly incorporates UAV dynamics, fleet size, and operational constraints, ensuring practical feasibility in real-world farming applications.

The UAV system is modeled as a linear time-varying (LTV) system, accounting for mass and moment of inertia variations due to chemical payload changes. The D$^2$OC-based control strategy, derived using Lagrangian mechanics, enables adaptive UAV operations, prioritizing critical areas while maintaining coverage efficiency. Additionally, the framework dynamically assigns UAV tasks based on a reference density map, ensuring balanced workload distribution and efficient spraying.

In summary, this work presents the following contributions: 
i) The spraying drone is modeled as a Linear Time-Varying (LTV) system to account for variations in mass and moments of inertia, capturing the dynamics relevant to agricultural spraying applications;
ii) The D$^2$OC framework is presented for decentralized multi-agent systems, achieving non-uniform area coverage and enabling efficient chemical spraying in smart farming scenarios;  
iii) Within the proposed scheme, the optimal control input is derived in closed form for LTV systems, facilitating spatially continuous navigation across the task domain. This approach distinguishes the present work from prior studies \cite{kabir2020ergodicity, kabir2021efficient, kabir2021wildlife, lee2022density}, which focus on path-planning over discrete sample points without explicit consideration of agent dynamics or coverage optimality;
iv) Simulations are conducted using both conventional and recent baselines to validate the proposed method, demonstrating superior weed-control performance and practical applicability.

\section{Problem Statement}
\textit{Notation:} Sets of real and natural numbers are denoted by $\mathbb{R}$ and $\mathbb{N}$, respectively. Further, $\mathbb{N}_0 = \mathbb{N} \cup {0}$. The symbols $\Vert \cdot \Vert$ and $^{\top}$, respectively, denote the Euclidean norm and the transpose operator. The variable $k \in \mathbb{N}_0$ is used to denote a discrete-time index. The identity matrix with dimension $n$ is denoted by $\mathbf{I}_n$. The symbol $\otimes$ denotes the Kronecker product. The notation $\operatorname{diag}(\cdot)$ represents a diagonal matrix with the specified components along its main diagonal.
\subsection{Problem Description} 
In smart farming applications, for a given chemical-spraying task to manage pests, fungi, diseases, and/or weeds, it is more desirable to cover farm areas with a given priority instead of uniform coverage. 
Compared to uniform coverage, priority-based coverage is more efficient at dealing with issues and more beneficial to the agricultural environment due to selective treatment.
The key idea to enabling such efficiency is to make a team of agricultural drones focus more on high-priority areas while spending less time on low-priority areas. 

\begin{figure}[!h]
	\centering
    \subfloat[Reference distribution map for damaged farm areas by pests, diseases, and weeds]{\includegraphics[scale=0.45]{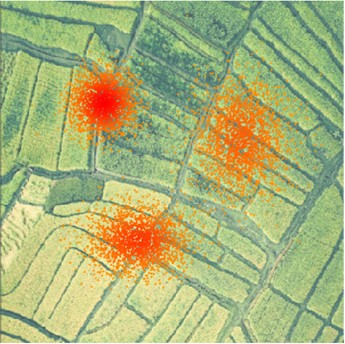}}\quad
    \subfloat[Schematic of density-driven optimal control]{\includegraphics[scale=0.45]{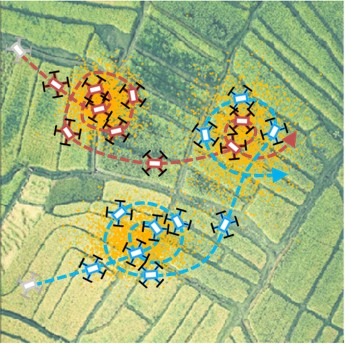}}

	\caption{Schematic of the proposed density-driven optimal control with the sample-point representation of the reference distribution.}
	\label{fig: robot_position_schematic}
\end{figure}

A conceptual figure is provided in Fig. \ref{fig: robot_position_schematic} to illustrate what we plan to achieve through this study. 
The red point clouds in Fig. \ref{fig: robot_position_schematic}(a) represent areas of illness on the farm, identified through satellite imagery and enhanced by AI analysis as in \cite{aitkenhead2003weed}. A team of spraying drones is then deployed to cover the farm based on this reference distribution map, applying solutions to the affected areas. The key challenge is \textit{how to control} the autonomous drones to prioritize higher-risk areas, ensuring that their trajectories align with the reference distribution, as depicted in Fig. \ref{fig: robot_position_schematic}(b). To address this challenge, Optimal Transport (OT) theory is utilized, which will be discussed in the following section.

\subsection{Preliminary: Optimal Transport}
Optimal transport is a research field focusing on studying optimal transportation or resource allocation \cite{monge1781memoire, villani2008optimal}. 
Analogous to measuring the distance between two points, such as Euclidean distance, optimal transport can be interpreted as measuring the distance between two probability density functions (PDFs).

Given two discrete distributions $\mu_1$ and $\mu_2$, containing $M$ and $N$ numbers of discrete points with some given weights, respectively, the squared 2-Wasserstein distance is defined by

{\small
\begin{align}
&\mathcal{W}_2^2(\mu_1, \mu_2) =\min_{\gamma_{ij}} \sum_{i,j}\gamma_{ij} \lVert y_i - q_j \rVert^2 \label{eqn: W_LP}\\
		&\text{subject to} \left \{
	\begin{aligned}\quad &\sum_{i=1}^{M}\gamma_{ij} = \beta_{j}, \, \forall j,\,\sum_{j=1}^{N}\gamma_{ij} = \alpha_{i}, \, \forall i, \,\\
        \quad& \sum_{i=1}^{M}\alpha_i = \sum_{j=1}^{N}\beta_j = 1,
	\end{aligned}\nonumber
	\right.
\end{align}
}where $y_i$ and $q_j$, respectively, are the discrete points included in $\mu_1$ and $\mu_2$, and $\alpha_i$ and $\beta_j$ are the \textit{non-negative} weights of discrete points $y_i$ and $q_j$, respectively. 
The coupling coefficient $\gamma_{ij}$ can be understood as a transportation plan, indicating the amount of weight taken out from the point $q_j$ and transported to the point $y_i$.

The first constraint ensures that the total weight transported from \( q_j \) to \( y_i\) equals \( \beta_j \), preserving the mass at each source point. Similarly, the second constraint enforces that the total weight received at \( y_i \) equals \( \alpha_i \), maintaining mass balance at each destination. The final constraint guarantees mass conservation, meaning no mass is created or lost—every unit of mass in \( \{\beta_j\} \) must be fully transported to \( \{\alpha_i\} \). The goal of optimal transport is then to determine the \textit{optimal transportation plan} \( \gamma_{ij}^{*} \), specifying how mass should be transported from \( q_j \) to \( y_i \) in the most efficient manner.

In this problem, the 2-Wasserstein distance \eqref{eqn: W_LP} forms a Linear Programming (LP) problem, and the solution for the optimal transport plan $\gamma_{ij}^{*}$ can be obtained by solving the above LP problem. For notational ease, the symbol $\mathcal{W}$ without the subscript will be used hereafter to denote the 2-Wasserstein distance.

\subsection{Problem Solving Approach}\label{sec:2.3}
The density distribution $\{q_j\}$ in \eqref{eqn: W_LP} is assumed to be available as a reference distribution.
Unlike the LP problem \eqref{eqn: W_LP}, where both $\{y_i\}$ and $\{q_j\}$ are known a priori, the agent trajectories $\{y_i\}$ are not yet determined in the density-driven optimal control problem.
Thus, our goal is to generate the multiple drones' trajectories $\{y_i\}$ such that the distribution formed by these trajectories will closely match the given reference distribution $\{q_j\}$. 
To clearly distinguish $\{y_i\}$ from $\{q_j\}$, we refer to $\{y_i\}$ and $\{q_j\}$ as \textit{agent}-points and \textit{sample}-points, respectively.

Here, we consider that the agent-points evolve sequentially in time, leading to their dependency on time. That is, the future agent positions depend on the previous and current agent positions. As a result, the agent-points will be explicitly expressed as $\{y^k\}$ with a discrete-time index $k$ instead of $\{y_i\}$. 
The left superscript $r$ denotes the index for the agent (or drone) in the multi-drone system.
Therefore, ${^ry^k}$ represents the position of the agent $r$ at discrete time $k$.

Given the total number of agents \( n_d \in \mathbb{N} \), let \( {}^rM \in \mathbb{N} \) denote the total number of agent-points along the trajectory of agent \( r \). This quantity is determined by dividing the agent’s nominal flight time by the discrete sampling interval, thereby implicitly accounting for the operation time of each drone. The weight assigned to an agent-point \( {^r y^k} \) at time step \( k \) is represented by \( {^r \alpha^k} \), which reflects the normalized energy consumption of agent \( r \) over the time interval from \( k-1 \) to \( k \).

Regarding the sample points \( \{q_j\}_{j=1}^{N} \), which define the priority map in smart farming applications to address issues such as pests, fungi, diseases, and weeds, the following conditions are imposed. Given \( N \in \mathbb{N} \) sample points representing the reference PDF, each sample point \( q_j \) is initially assigned a uniform weight, i.e., \( {}^r\beta_j^{k=0} = \frac{1}{N} \). The regional priority is then represented by distributing more sample points in higher-priority areas. 

The weight \( {}^r\beta^k_j \) evolves over time, decreasing as agents explore the domain. This reduction ensures that previously visited or nearby sample-points receive lower priority, which must be captured in the time-varying weight \( {}^r\beta^k_j \). Therefore, a suitable update rule must be established to adjust \( {}^r\beta^k_j \) based on the agents' traversal history. Additionally, due to the decentralized control setup, the weight \( {}^r\beta^k_j \) may vary between agents, necessitating the inclusion of the agent index \( r \).

Considering agent constraints, including agent dynamics, the number of agents, and energy limitations, the objective of the controller is to generate agent-points such that
\(
\sum_{r=1}^{n_d}\sum_{k=1}^{{}^rM}\sum_{j=1}^{N}{}^r\gamma_{kj}\Vert {}^r y^k - q_j\Vert
\)
is minimized while satisfying the OT constraints in \eqref{eqn: W_LP}, where \( {}^{r}\gamma_{kj} \) represents the transportation from the sample point \( q_j \) to the agent-point \( {}^r y^{k} \).
Achieving this goal minimizes the Wasserstein distance between the density formed by the multi-agent trajectories and the reference map.

\section{Dynamics of the Spraying Drone}
For an agricultural drone used for spraying chemicals to address agricultural issues, the mass and moment of inertia change over time as the drone dispenses the solution onto the farm. This section presents how the spraying drone system can be modeled as a linear time-varying (LTV) system to account for the gradual reduction in the sprayed material.
 
\begin{figure}[!h]
      \centering
      \includegraphics[scale=0.48]{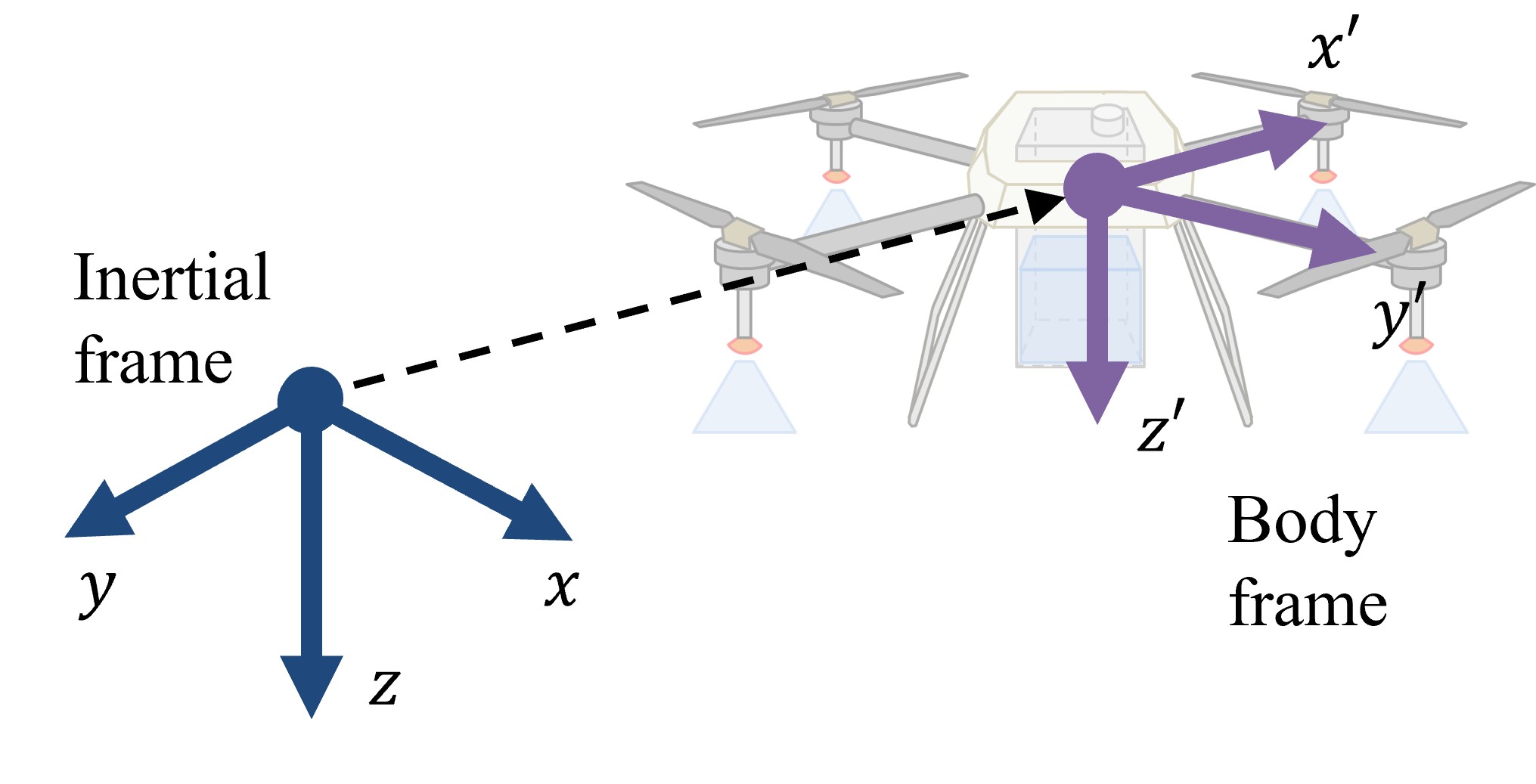}
      \caption{Representation of the inertial frame and the body frame of the drone.}
      \label{fig: QR_coord}
  \end{figure}

Fig. \ref{fig: QR_coord} illustrates the inertial and body frames of the drone.  
The three-dimensional position vector of the drone in the inertial frame is given by $[x\ y\ z]^\top$, while the Z-Y-X Euler angles vector, representing the rotation from the inertial frame to the body frame, is defined as $[\phi\ \theta\ \psi]^\top$, corresponding to roll, pitch, and yaw angles, respectively.  

The translational and angular velocities of the drone along the $x'$-, $y'$-, and $z'$-axes in the body frame are denoted by $[u\ v\ w]^\top$ and $[p\ q\ r]^\top$, respectively. Under the assumption of negligible wind disturbances and small angle approximations ($\phi,\ \theta,\ \psi \approx 0$), the linearized drone dynamics can be expressed in terms of the state vector  
$ \mathsf{x} = [\phi\ \theta\ \psi\ p\ q\ r\ u\ v\ w\ x\ y\ z]^\top $  
and the control input vector  
$ u = [f_t\ \tau_x\ \tau_y\ \tau_z]^\top.$ 

Following the derivation in \cite{sabatino2015quadrotor}, the system dynamics are given by  
$ \dot{\mathsf{x}} = A_c\mathsf{x} + B_c u, \label{eqn: cont_QR_dynamics} $  
where $A_c$ and $B_c$ are the system matrices for the linearized drone dynamics as presented in \cite{sabatino2015quadrotor}.

The moments of inertia about $x',\ y',\ $and $z'$ axes in the body frame are denoted by $I_{x'},\ I_{y'},\ $and $I_{z'}$, the symbol $m$ represents the mass of the drone, and the gravity of Earth is denoted by $g$.  

The discrete-time dynamics of the linearized drone are then obtained by 
\begin{align}
    \mathsf{x}^{k+1} = \underbrace{(\mathbf{I}_{12}+\Delta T\cdot A_c)}_{A_k}\mathsf{x}^{k} + \underbrace{\Delta T\cdot B_c}_{B_k}u^k,\label{eqn: difference LTV}
\end{align}
where $\Delta T$ represents the discrete-time step size.

As the spraying drones disperse chemical solutions over the farm area, the system's mass and moments of inertia, denoted as \( m \), \( I_{x'} \), \( I_{y'} \), and \( I_{z'} \), continuously change, leading to the time-varying structure of \eqref{eqn: difference LTV}. To simplify the analysis, the following assumptions are introduced.
\begin{assumption}\label{assumption}
    The spray tank is assumed to have the shape of a square prism. Furthermore, the drone is assumed to fly at a relatively low speed, which minimizes fluctuations of the solution within the spray tank. Under these assumptions and applying the small-angle approximation ($\phi, \ \theta \approx 0$), the solution will retain its square prism shape. Additionally, assuming the spray tank is symmetrically mounted on the drone and the solution mass is relatively small compared to the drone, the center of mass of the combined system (drone and solution tank) is approximated to coincide with the center of mass of the drone.
\end{assumption}

  \begin{figure}[!h]
      \centering
      \includegraphics[scale=0.4]{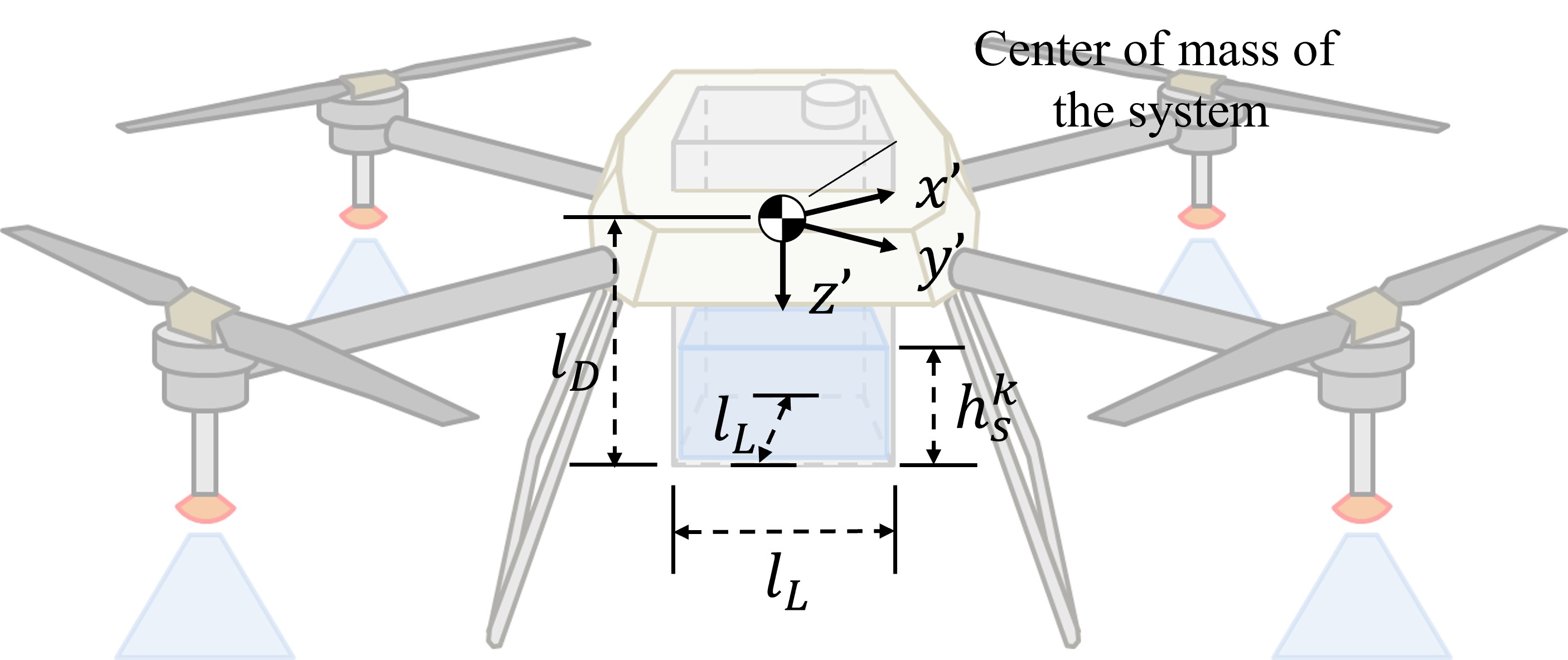}
      \caption{Illustration of the notations related to the solution and the spray tank.}
      \label{fig: QR_Spraying}
  \end{figure}
Fig. \ref{fig: QR_Spraying} illustrates the spraying drone system composed of a drone and a spray tank. The width and length of the spray tank are denoted by $l_L$ and the distance between the center of mass and the bottom of the spray tank is represented by $l_D$. The height of the solution at time $k$ is indicated by $h_{s}^{k}$ and the spraying rate (volumetric flow rate) at time $k$ by $Q_{s}^{k}$. Then, the height of the solution, $h^k_{s}$, and the mass of the system, $m^{k}$, evolve over time by
\begin{align}
   h_{s}^{k+1} = h_{s}^{k}-\Delta T\frac{Q_{s}^{k}}{l_L^2},\quad
   \ m^k_s = \rho_s l_L^2 h_{s}^{k},\quad
   m^{k} = m_d+m^k_s, \label{eqn: m^k}
\end{align}
where $m_d$ is the mass of the drone, $m^k_s$ is the mass of the solution at time $k$, and $\rho_s$ indicates the density of the solution. Utilizing these two equations, the moments of inertia of the system about $x'$-, $y'$-, $z'$- axes, $I_{x'}^{k},\ I_{y'}^{k},\ I_{z'}^{k}$, are calculated in the following proposition.

\begin{proposition}\label{prop: time-varying varibles}
Consider a spraying drone system (Fig. \ref{fig: QR_Spraying}) obeying Assumption \ref{assumption}. This system has time-varying parameters for the height of the solution and the mass as in \eqref{eqn: m^k}.
Then, the moments of inertia of the system about $x'$-, $y'$-, and $z'$-axes, denoted by $I_{x'}^{k}$, $I_{y'}^{k}$, $I_{z'}^{k}$, are calculated by
\begin{align}
   \begin{aligned} &I_{x'}^{k}=I_{d,x}+m_{s}^{k}\left(l_D-\frac{h_{s}^{k}}{2}\right)^2+\frac{m_{s}^{k}}{12}\left(l_L^2+\left(h_{s}^k\right)^2\right),
   \\& I_{y'}^{k}=I_{d,y}+m_{s}^{k}\left(l_D-\frac{h_{s}^{k}}{2}\right)^2+\frac{m_{s}^{k}}{12}\left(l_L^2+\left(h_{s}^k\right)^2\right),
   \\& I_{z'}^{k} = I_{d,z}+\frac{m_{s}^k}{6}l_L^2,
   \end{aligned}\label{eqn: MOI}
\end{align}
where $I_{d,x}$, $I_{d,y}$, and $I_{d,z}$ are the moments of inertia of the drone about the $x$, $y$, and $z$ axes, respectively, measured at the center of mass of the system. 
Consequently, the dynamics of the linearized spraying drone are obtained by
\begin{align}
    \mathsf{x}^{k+1} = A_k \mathsf{x}^k + B_k u^k,\label{eqn: LTV_SprayingDrone}
\end{align}
where
\begin{align*}
    &A_k = \begin{bmatrix}
        \mathbf{I}_3&\Delta T\cdot\mathbf{I}_3&\mathbf{0}&\mathbf{0}\\
        \mathbf{0}&\mathbf{I}_3&\mathbf{0}&\mathbf{0}\\
        \mathbf{G} &\mathbf{0}& \mathbf{I}_3 & \mathbf{0}\\
        \mathbf{0}&\mathbf{0}&\Delta T\cdot\mathbf{I}_3 & \mathbf{I}_3
    \end{bmatrix}, B_k = \begin{bmatrix}
        \mathbf{0}_{3\times1} & \mathbf{0}_{3\times3}\\
        \mathbf{0}_{3\times1} & \mathbf{I}^k_{xyz}\\
        \mathbf{M}^k&\mathbf{0}_{3\times3}\\\mathbf{0}_{3\times1} & \mathbf{0}_{3\times3}
    \end{bmatrix},
    \\&\mathbf{G}=\Delta T \begin{bmatrix}
        0&-g&0\\
        g&0&0\\
        0&0&0
    \end{bmatrix},\ \mathbf{I}^k_{xyz} = \Delta T \begin{bmatrix}
        1/I^k_{x'}&0&0\\
        0&1/I^k_{y'}&0\\
        0&0&1/I^k_{z'}
    \end{bmatrix},
    \\&\mathbf{M}^k = \Delta T \begin{bmatrix}
        0\\0\\-1/m^k
    \end{bmatrix},
\end{align*} with $g$ denoting gravitational acceleration, $m^k$ denoting the mass of the system, as defined in \eqref{eqn: m^k}, and $\Delta T$ denoting the discrete-time step size.
\end{proposition}

\begin{figure}[!h]
    \centering
    \includegraphics[width=0.6\linewidth]{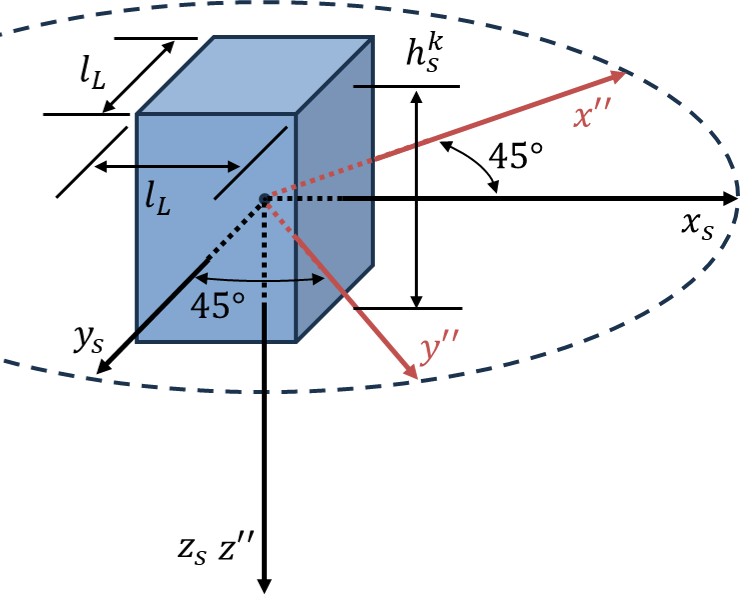}
    \caption{Moments of inertia of the solution in the spray tank.}
    \label{fig: Inertia_spraytank}
\end{figure}
\begin{proof}
Fig. \ref{fig: Inertia_spraytank} presents the dimensions of the solution in the spray tank and the axes about which the moments of inertia are calculated. The $x_s$-$y_s$-$z_s$ coordinate frame is located at the center of mass of the solution. The $x''$-$y''$-$z''$ coordinate frame is rotated by 45 degrees relative to the $x_s$-$y_s$-$z_s$ coordinate frame. The $x''$ and $y''$ axes are parallel to the $x'$ and $y'$ axes in Fig. \ref{fig: QR_Spraying}, respectively.

The moments of inertia of the solution about the $x_s$-, $y_s$-, and $z_s$-axes, $I_{s,x_s}^{k}$, $I_{s,y_s}^{k}$, $I_{s,z_s}^{k}$, are calculated by
\begin{align*}
&\begin{aligned}
    I_{s,x_s}^{k} &= \int_{-\frac{l_L}{2}}^{\frac{l_L}{2}}\int_{-\frac{l_L}{2}}^{\frac{l_L}{2}}\int_{-\frac{h_s^k}{2}}^{\frac{h_s^k}{2}} (y^2+z^2)\rho_s\ dzdydx\\
    &= \frac{m^k_s}{12}\left(\left(h^k_s\right)^2+l_L^2\right),\\
    I_{s,y_s}^{k} &= \int_{-\frac{l_L}{2}}^{\frac{l_L}{2}}\int_{-\frac{l_L}{2}}^{\frac{l_L}{2}}\int_{-\frac{h_s^k}{2}}^{\frac{h_s^k}{2}} (z^2+x^2)\rho_s\ dzdydx \\
    &= \frac{m^k_s}{12}\left(\left(h^k_s\right)^2+l_L^2\right),\\
    I_{s,z_s}^{k} &= \int_{\frac{l_W}{2}}^{\frac{l_W}{2}}\int_{\frac{l_L}{2}}^{\frac{l_L}{2}}\int_{\frac{h_s^k}{2}}^{\frac{h_s^k}{2}} (x^2+y^2)\rho_s\ dzdydx = \frac{m^k_s}{6}l_L^2,
\end{aligned}
\end{align*}
where $h_s^{k}$ and $m_s^{k}$ are given in \eqref{eqn: m^k}.

As the axes $x_s$, $y_s$, and $z_s$ are the principal axes of the solution, the products of inertia are zero, and the inertia tensor about $x_s$-, $y_s$-, and $z_s$-axes, $I^k_s$, is obtained by
$
    I_s^k = \frac{m^k_s}{12}\times 
    \operatorname{diag}\left(\left[\left(h^k_s\right)^2+l_L^2,  \left(h^k_s\right)^2+l_L^2, 2l_L^2\right]\right).
$

Since the $x''$, $y''$, and $z''$ frame is rotated by 45 degrees about the $z_s$ axis from the $x_s$, $y_s$, and $z_s$ frame, the inertia tensor about the $x''$, $y''$, and $z''$ frame, $I^k_{s''}$, is evaluated using a rotation matrix by
\begin{align}
    I^k_{s''} &= \begin{bmatrix}
        \cos{45^\circ}&-\sin{45^\circ}&0
        \\\sin{45^\circ}&\cos{45^\circ}&0
        \\0&0&1
    \end{bmatrix}I^k_s
    \begin{bmatrix}
        \cos{45^\circ} & \sin{45^\circ} & 0
        \\-\sin{45^\circ} & \cos{45^\circ} & 0
        \\0 & 0 & 1
    \end{bmatrix}\nonumber
    \\&=\frac{m^k_s}{12}\begin{bmatrix}
        \left(h^k_s\right)^2+l_L^2 & 0 & 0
        \\0&\left(h^k_s\right)^2+l_L^2 &0
        \\0&0&2l_L^2
    \end{bmatrix}.\nonumber
\end{align}
Applying the parallel axis theorem, the total moments of inertia of the system (drone and solution) about the \( x' \)-, \( y' \)-, and \( z' \)-axes are given by \eqref{eqn: MOI}.

By incorporating the time-varying parameters \( m^k \), \( I_{x'}^{k} \), \( I_{y'}^{k} \), and \( I_{z'}^{k} \) into \eqref{eqn: difference LTV}, the LTV dynamics of the spraying drone system are given by \eqref{eqn: LTV_SprayingDrone}.
\end{proof}
\begin{remark}
    The LTV representation used in this work is justified by the standard small-angle approximation for multirotor UAVs operating in nominal flight conditions. This approximation enables the optimal control problem in Section~IV to be expressed with linear dynamics, allowing a closed-form control law to be derived without iterative nonlinear optimization. In contrast, nonlinear formulations (e.g., NMPC or backstepping) require repeated numerical computations or complex gain tuning, reducing real-time applicability. The LTV representation therefore provides a practical balance between model fidelity and computational efficiency.
\end{remark}

\section{Density-Driven Optimal Control (D$^2$OC)}\label{sec: D2OC}
Building upon the LTV system structure, this section outlines the formal procedure for Density-Driven Optimal Control (D$^2$OC), which is composed of three distinct stages: Stage A (Optimal Control), Stage B (Weight Update), and Stage C (Weight Sharing).

Prior to deployment, each agent is provided with the total number of agent-points, $\sum_{r=1}^{n_d} {}^rM$, the positions of the sample points $\{q_j\}_{j=1}^N$, and their initial weights $\{\beta_j^0\}_{j=1}^N$, which collectively constitute the agent's initial local weight data. During operation, each agent updates this data at every time step based on its own coverage and information exchanged with neighboring agents, as detailed below. In \textbf{Stage A}, each drone selects a set of local sample-points and computes the optimal control input to minimize the Wasserstein distance. After executing the control input, the sample-point weights are updated in \textbf{Stage B}, where the weights of nearby sample-points are transferred to the drone, increasing the likelihood of visiting higher-weight regions in the next time step. For clarity, the agent index \( r \) is omitted in Stages A and B, as these stages are commonly applied to each agent independently.  
\textbf{Stage C} enables weight sharing among agents within their communication range, promoting decentralized control. This improves coordination and reduces redundant coverage in multi-agent systems.

These three stages are executed sequentially at each discrete time step, forming the complete D$^2$OC framework.


\subsection{Stage A. Optimal Control Stage} 
The optimal control stage consists of two steps: first, the selection of local sample-points, and second, the computation of the optimal control input to minimize the Wasserstein distance.

The local sample-points selection step is necessary to identify a subset of sample-points from the entire set, reducing the computational complexity of the optimization. Directly solving the Wasserstein distance minimization problem for all sample-points and agent-points simultaneously is computationally intractable, especially for large-scale missions, due to its highly non-convex nature. As the number of sample points increases, the number of possible transport plans grows exponentially, significantly amplifying both the computational complexity and the difficulty of solving the problem.

By first selecting local sample-points, the subsequent optimal control step can then determine the control input that minimizes the Wasserstein distance between the current agent-point and the selected local sample-points efficiently. This hierarchical approach ensures that the optimization remains feasible while still maintaining the accuracy and effectiveness of the coverage strategy.

\begin{figure}[!h]
    \centering
    \includegraphics[scale=0.6]{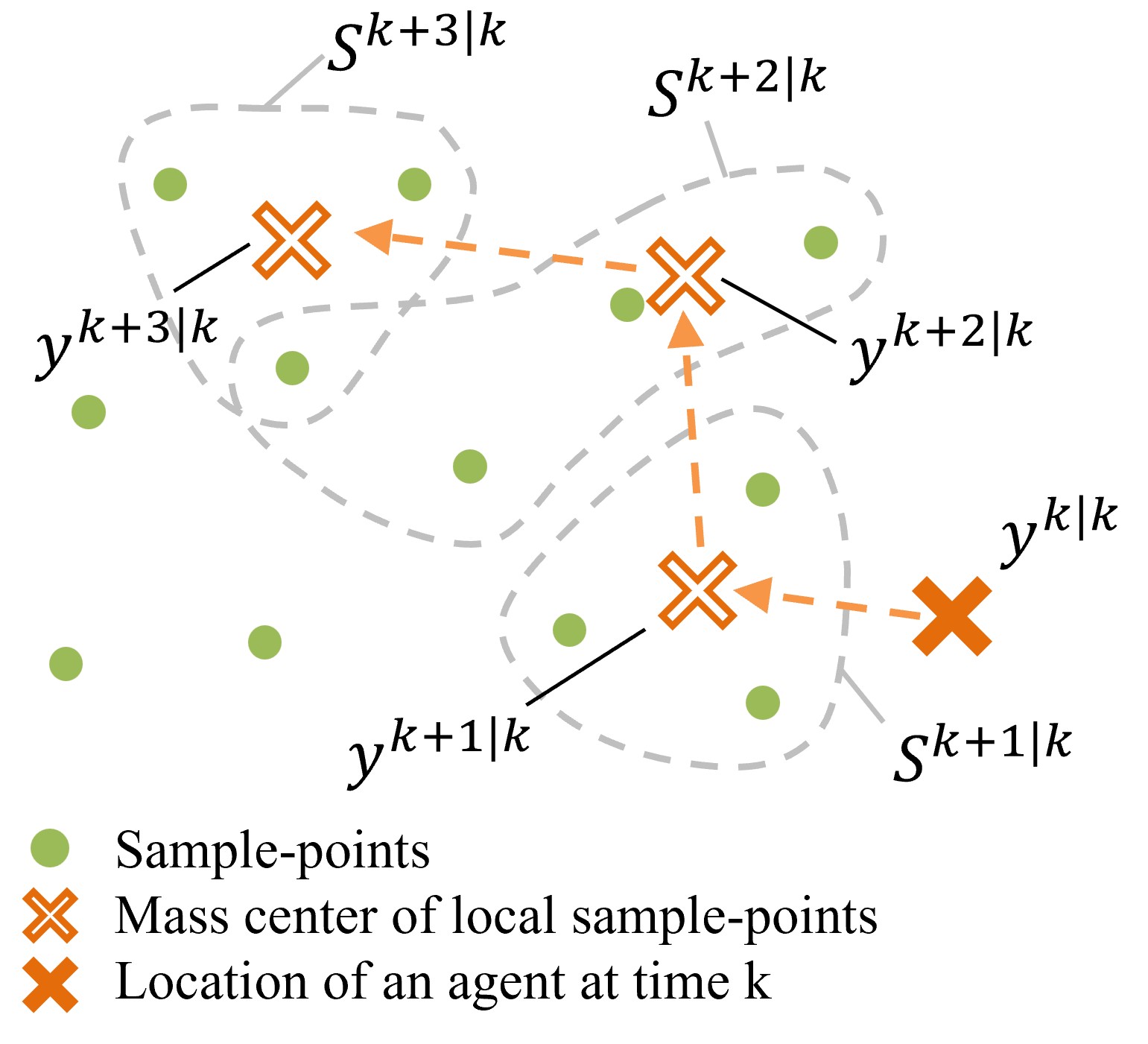}
    \caption{Stage A - prediction of the local sample-points over the finite prediction horizon $T$ = 3. The set of predicted local sample-points at time $k+i$ when the current time is $k$ is denoted by $S^{k+i|k}$, and $y^{k+i|k}$ is the mass center of the local sample-points contained in the $S^{k+i|k}$.}
    \label{fig: Pred_Loc_SP}
\end{figure}

Fig. \ref{fig: Pred_Loc_SP} illustrates the process of selecting local sample-points. At time \(k\), for a given prediction horizon \(T\), a total of \(T\) subsets of local sample-points are determined: \(\mathcal{S}^{k+1|k}, \mathcal{S}^{k+2|k}, \dots, \mathcal{S}^{k+T|k}\). These subsets serve as reference sets for the agent to track throughout the prediction horizon.

Specifically, the subset \( \mathcal{S}^{k+i|k} \), for \(i = 1, 2, \dots, T\), is determined based on the weight-normalized (wn) distance, \( d_{\text{wn}}(j, y^{k+i-1|k}) \), defined as 
\begin{align}
    d_{\text{wn}}(j, y^{k+i-1|k}) = \dfrac{ \| q_{j} - y^{k+i-1|k} \| }{ \beta^k_{j} - \sum_{l=1}^{i-1} \beta^{k+l|k}_j }, \label{eqn: weight-normalized distance}
\end{align}
where \( q_{j} \) is the location of sample-point \(j\) and \( \beta^k_{j} \) represents the weight of sample-point \(j\) at time \(k\). The term \( \beta^{k+l|k}_j \) corresponds to the weight of sample-point \(j\) in the subset \( \mathcal{S}^{k+l|k} \). The term \( y^{k+i-1|k} \) denotes the target location of the agent at the \((i-1)\)-th prediction step and is defined as
\begin{align}
    y^{k+i-1|k} = \dfrac{ \sum_{j \in \mathcal{S}^{k+i-1|k}} \beta_j^{k+i-1|k} q_{j} }{ \sum_{j \in \mathcal{S}^{k+i-1|k}} \beta_j^{k+i-1|k} }, \quad i=2,...,T,\label{eqn: y - center of mass}
\end{align}
where $y^{k|k}=y^k$ denotes the agent's position at time $k$.

This term \( y^{k+i-1|k} \) for $i=2,...,T$ represents the center of mass of the subset \( \mathcal{S}^{k+i-1|k} \). The weight-normalized distance \( d_{\text{wn}} \) serves as a prioritization measure, guiding the agent to favor sample-points that are both closer to the agent and have higher weights, thus enhancing the efficiency of control input synthesis.

The selection of local sample-points proceeds in ascending order of weight-normalized distance \( d_{\text{wn}} \), with each successive sample-point chosen based on the next smallest value of \( d_{\text{wn}} \). During this process, the weight of sample-point \( j \) in \( \mathcal{S}^{k+i|k} \), denoted \( \beta^{k+i|k}_j \), is updated as follows:
\begin{align*}
\beta^{k+i|k}_j = 
\begin{cases} 
\beta^k_{j} - \displaystyle\sum_{l=1}^{i-1} \beta^{k+l|k}_j & \text{if } \beta^k_{j} - \displaystyle\sum_{l=1}^{i-1} \beta^{k+l|k}_j > \alpha^{k+i|k}_{rem}, \\
\alpha^{k+i|k}_{rem} & \text{otherwise}.
\end{cases}
\end{align*}
Here, \( \alpha^{k+i|k}_{rem} \) represents the remaining weight of the subset \( \mathcal{S}^{k+i|k} \), which initially equals \( \alpha^{k+i} \). As each sample-point \( j \) is added to \( \mathcal{S}^{k+i|k} \), the remaining weight is updated as $
\alpha^{k+i|k}_{rem} = \alpha^{k+i|k}_{rem} - \beta^{k+i|k}_j.$

This process continues until \( \alpha^{k+i|k}_{rem} \) reaches zero, indicating that all of the remaining weight has been allocated.

The formal process to select local sample-points is provided in Algorithm \ref{algorithm:1}.
\begin{algorithm}[h!]
	\caption{Local Sample-points Selection Process}\label{algorithm:1}
		\hspace*{\algorithmicindent} \textbf{Input:}  $\alpha^{k+1},\ \alpha^{k+2},\ ...,\ \alpha^{k+T},\  \beta_j^k,\ q_j,\ \forall j$
  \begin{algorithmic}[1]	
  \For{$i = 1,2, ..., T$}
\State $\mathcal{S}^{k+i|k} \gets \O$
\State $\alpha^{k+i|k}_{rem} \gets \alpha^{k+i}$
\If{$i=1$}
\State $y^{k+i-1|k} \gets y^k$
\Else
\State Calculate $y^{k+i-1|k}$ from \eqref{eqn: y - center of mass}
\EndIf
\While{$\alpha^{k+i|k}_{rem}\neq0$} 
    \State $j_{min} \gets \underset{j}{\operatorname{argmin}} \ d_{\text{wn}}(j,y^{k+i-1|k})$
    \State $\mathcal{S}^{k+i|k} \gets \mathcal{S}^{k+i|k}\cup \{j_{min}\}$
    \If{$\beta^k_{j_{min}}-\sum_{l=1}^{i-1} \beta^{k+l|k}_{j_{min}}>\alpha^{k+i|k}_{rem}$}
    \State $\beta^{k+i|k}_{j_{min}} \gets \alpha^{k+i|k}_{rem}$
    \Else
    \State $\beta^{k+i|k}_{j_{min}} \gets \beta^k_{j_{min}}-\sum_{l=1}^{i-1} \beta^{k+l|k}_{j_{min}}$
    \EndIf
    \State $\alpha^{k+i|k}_{rem} \gets \alpha^{k+i|k}_{rem} - \beta_{j_{min}}^{k+i|k}$
		\EndWhile
\EndFor	\end{algorithmic}
\end{algorithm}

\begin{remark}
The target location of the agent at the \(i\)-th step ahead of time \(k\) is determined by minimizing the Wasserstein distance between the subset \( \mathcal{S}^{k+i|k} \) and the predicted agent location at time \(k+i\), which is yet to be computed. This minimization is expressed as \( \displaystyle\min_{\gamma_j, y^{k+i|k}} \sum_{j \in \mathcal{S}^{k+i|k}} \gamma_j \| y^{k+i|k} - q_j \|^2 \), where \( \gamma_j \) denotes the transport plan that maps the sample points in \( \mathcal{S}^{k+i|k} \) to the target location \( y^{k+i|k} \). Since the weights of these points are transported to a single agent's location, the transportation plan \( \gamma_j \) is simply replaced by \( \beta^{k+i|k}_j \). Then, this Wasserstein distance is minimized when \( y^{k+i|k} \) coincides with the center of mass of \( \mathcal{S}^{k+i|k} \), which is given by \( y^{k+i|k} = \frac{\sum_{j \in \mathcal{S}^{k+i|k}} \beta_j^{k+i|k} q_j}{\sum_{j \in \mathcal{S}^{k+i|k}} \beta_j^{k+i|k}} \). This explains why \( y^{k+i|k} \) is defined as the mass center of \( \mathcal{S}^{k+i|k} \), thereby justifying its structure as shown in \eqref{eqn: y - center of mass} for the target location.
\end{remark}

The next step is to obtain the optimal control input for the spraying drone system using these subsets obtained at time $k$.
For the spraying drone system, consider a discrete-time LTV system dynamics given by 
\begin{equation}
    \begin{aligned}
        &\mathsf{x}^{k+1} = A_k\mathsf{x}^k + B_ku^k, \qquad
        y^k = C\mathsf{x}^k,
    \end{aligned}\label{eqn: LTV system}
\end{equation}
where the state vector and system matrices are already defined for the spraying drone system in \eqref{eqn: difference LTV}--\eqref{eqn: MOI}.
Note that both \( A_k \) and \( B_k \) are time-varying due to the variation in mass, which in turn affects the moment of inertia.

The following optimization problem is then formulated for D$^2$OC in association with the Wasserstein distance using the subset of local-sample points as follows.
\begin{align}
&\begin{aligned}
	&\min_{u} J = \Phi(\mathsf{x}^{k+T})
 +\\
 &\sum_{i=k}^{k+T-1}\left(\dfrac{1}{2}(\mathcal{W}^{i|k})^2 + \dfrac{1}{2}(\mathsf{x}^{i})^{\top}Q\mathsf{x}^{i} +  \dfrac{1}{2}(u^{i})^{\top}Ru^{i}\right)
 \end{aligned}\label{eqn: minimization problem}\\
 &\text{subject to} \quad \mathsf{x}^{i+1} = A_i\mathsf{x}^i + B_i u^i,\nonumber
\end{align}
where
\(
(\mathcal{W}^{i|k})^2 = \min_{\gamma_{j}} \sum_{j \in \mathcal{S}^{i|k}} \gamma_j \| C\mathsf{x}^{i} - q_j \|^2, \quad \text{for } i \neq k, \quad \text{and} \quad (\mathcal{W}^{i|k})^2 = 0 \quad \text{for } i = k.
\)
Also, 
\(
\Phi(\mathsf{x}^{k+T}) = \frac{1}{2} (\mathcal{W}^{k+T|k})^2 + \frac{1}{2} (\mathsf{x}^{k+T})^{\top} Q \mathsf{x}^{k+T}.
\)

In \eqref{eqn: minimization problem}, $\Phi(\mathsf{x}^{k+T})$ represents the terminal cost and promotes terminal-state regulation, resulting in a more tightly controlled final behavior. The term $\mathcal{W}^{i|k}$ denotes the Wasserstein distance between the local-sample points in $\mathcal{S}^{i|k}$ and the agent-point $y^i$.
The third and fourth terms in the objective function are to penalize the state vector and input vector with a positive semi-definite matrix $Q\in \mathbb{R}^{n\times n}$ and a positive definite matrix $R\in \mathbb{R}^{m\times m}$.
Then, the following theorem is proposed to provide the optimal solution of \eqref{eqn: minimization problem} to achieve D$^2$OC.

\begin{theorem}
     For the optimization problem \eqref{eqn: minimization problem} associated with the LTV system \eqref{eqn: LTV system} to model the time-varying nature of the spraying drones, the optimal control input $u^{\star|k}$ that minimizes the objective function \eqref{eqn: minimization problem} for D$^2$OC is derived by 
    \begin{align}
    \begin{aligned}
        &u^{\star|k} = \begin{bmatrix}
        \mathbf{I}_m & \mathbf{0} & \cdots & \mathbf{0}
        \end{bmatrix}\bar{u}^k \text{ with } \\ &\bar{u}^k = -\overline{E}E_{23}^{\top}E_{12}^{-1}F_{1} +\overline{E}E_{23}^{\top}E_{12}^{-1}E_{11}(E_{12}^{-1})^\top F_{2},
        &
        \end{aligned}
        \label{eqn: optimal_u}
    \end{align}
    where
    {\allowdisplaybreaks
    \begin{align*}
    &E_{11} = Q \otimes \mathbf{I}_{T} \,+ \\
    \qquad&\operatorname{diag}\left(\left[(\sum\limits_{j \in \mathcal{S}^{k+1|k}} \gamma_j)C^{\top}C, \ldots, (\sum\limits_{j \in \mathcal{S}^{k+T|k}} \gamma_j)C^{\top}C\right]\right),
    \nonumber
    \\&E_{12} = E_{21}^{\top} = 
    {\small
    \begin{bmatrix}
    -\mathbf{I}_n & A_{k+1}^{\top} & \mathbf{0}_{} & \cdots & \mathbf{0}_{}
    \\[0.1em]\mathbf{0}_{} & -\mathbf{I}_n & A_{k+2}^{\top} &  & \vdots
    \\[0.1em]\mathbf{0}_{} & \mathbf{0}_{} & -\mathbf{I}_n & \ddots & \mathbf{0}_{}
    \\[0.1em]\vdots & \vdots & & \ddots & A_{k+T-1}^{\top} 
    \\[0.7em]\mathbf{0}_{} & \mathbf{0}_{} & \dots & \mathbf{0}_{} & -\mathbf{I}_n  
    \end{bmatrix}}, \,\\
    &E_{23} = E_{32}^{\top} = 
    \operatorname{diag}\left(\left[ B_k, \ldots, B_{k+T-1}\right]\right),
    \, \\
    &E_{33} = R \otimes \mathbf{I}_{T}, \,\\
    &\overline{E}= (E_{33}+E_{23}^\top
    E_{12}^{-1}E_{11}{(E_{12}^{-1})^\top}E_{23})^{-1},\nonumber\\
    &F_1 = 
    \begin{bmatrix}
    (\sum\limits_{j \in \mathcal{S}^{k+1|k}} \gamma_j)C^{\top}\overline{q}^{k+1|k}\\
    (\sum\limits_{j \in \mathcal{S}^{k+2|k}} \gamma_j)C^{\top}\overline{q}^{k+2|k}\\
    \vdots\\
    (\sum\limits_{j \in \mathcal{S}^{k+T|k}} \gamma_j)C^{\top}\overline{q}^{k+T|k}
    \end{bmatrix},\,\,
    \overline{q}^{i|k}=\dfrac{\sum\limits_{j \in \mathcal{S}^{i|k}} \gamma_j q_{j}}{\sum\limits_{j \in \mathcal{S}^{i|k}} \gamma_j}, \nonumber\\
    &F_2 = 
    \left[
    (-A\mathsf{x}^k)^{\top}, \mathbf{0}_{}^{\top}, \ldots, \mathbf{0}_{}^{\top}
    \right]^{\top}.
    \end{align*}
    }
\end{theorem}
\begin{proof}
Given the optimization problem \eqref{eqn: minimization problem}, the Lagrangian is formulated by
\begin{equation}
    \begin{aligned}
         \mathcal{L} = &\Phi(\mathsf{x}^{k+T})+\sum_{i=k}^{k+T-1}\left(\dfrac{1}{2}(\mathcal{W}^{i|k})^2 + \dfrac{1}{2}(u^{i})^{\top}Ru^{i}+\right.\\
         &\left.\dfrac{1}{2}(\mathsf{x}^{i})^{\top}Q\mathsf{x}^{i}+(\lambda^{i+1})^{\top}(A_i \mathsf{x}^{i}+B_i u^{i}-\mathsf{x}^{i+1})\right),
     \end{aligned}\label{eqn: Lagrangian}
    \end{equation}
where $\lambda^{i+1}$ is a co-state vector.

The necessary equations and the boundary condition for the control input to be optimal are given by
{\allowdisplaybreaks
\begin{subequations}\label{eqn: pontryagin}
    \begin{align}
    	\dfrac{\partial \mathcal{L}}{\partial \mathsf{x}^{i}} = 0 &= 
        \dfrac{1}{2}\dfrac{\partial\mathcal(\mathcal{W}^{i|k})^2}{\partial \mathsf{x}^{i}} + Q\mathsf{x}^{i} + A_i^{\top}\lambda^{i+1}-\lambda^{i}\\
        &=\sum_{j \in \mathcal{S}^{i|k}} \gamma_j C^{\top}(C\mathsf{x}^{i}-q_j)+ Q\mathsf{x}^{i} + A_i^{\top}\lambda^{i+1}-\lambda^{i}\nonumber\\
        &\begin{aligned}
        &= \{(\sum_{j \in \mathcal{S}^{i|k}} \gamma_j)C^{\top}C+ Q\}\mathsf{x}^{i}
        -(\sum_{j \in \mathcal{S}^{i|k}} \gamma_j)C^{\top}\overline{q}^{i|k} + \\
        & \qquad A_i^{\top}\lambda^{i+1}-\lambda^{i}, 
        \end{aligned}
        \nonumber
        \\\dfrac{\partial \mathcal{L}}{\partial \lambda^{i'+1}} &= 0 = A_{i'}\mathsf{x}^{i'} + B_{i'}u^{i'} - \mathsf{x}^{i'+1}, \\
       \dfrac{\partial \mathcal{L}}{\partial u^{i'}} &=0 = Ru^{i'} + B_{i'}^{\top}\lambda^{i'+1} ,\\
       \dfrac{\partial \mathcal{L}}{\partial \mathsf{x}^{k+T}} &= 0 = 
         \{(\sum_{j \in \mathcal{S}^{k+T|k}} \gamma_j)C^{\top}C+ Q\}\mathsf{x}^{k+T}\nonumber\\
         & \qquad -(\sum_{j \in \mathcal{S}^{k+T|k}} \gamma_j)C^{\top}\overline{q}^{k+T|k} - \lambda^{k+T},
    \end{align}
    \end{subequations}
    }where $i = k+1,\ k+2,\ ...\ ,\ k+T-1$ and $i' = k,\ k+1,\ ...\ ,\ k+T-1$. By rearranging \eqref{eqn: pontryagin}, the resulting equation in matrix form is obtained by
    \begin{align}
        &E\begin{bmatrix}
        \Bar{\mathsf{x}}^{k+1} \\
        \Bar{\lambda}^{k+1} \\
        \Bar{u}^{k}
        \end{bmatrix} =
        \begin{bmatrix}
        E_{11} & E_{12} & \mathbf{0}_{} \\
        E_{21} & \mathbf{0} & E_{23}\\
        \mathbf{0} & E_{32} & E_{33}
        \end{bmatrix} 
        \begin{bmatrix}
        \Bar{\mathsf{x}}^{k+1} \\
        \Bar{\lambda}^{k+1} \\
        \Bar{u}^k
        \end{bmatrix} =
        \begin{bmatrix}
        F_{1} \\
        F_{2} \\
        \mathbf{0}
        \end{bmatrix}\nonumber\\
        &\rightarrow \begin{bmatrix}
        \Bar{\mathsf{x}}^{k+1} \\
        \Bar{\lambda}^{k+1} \\
        \Bar{u}^{k}
        \end{bmatrix} =
        E^{-1}
        \begin{bmatrix}  
        F_{1} \\
        F_{2} \\
        \mathbf{0}
        \end{bmatrix},\label{eqn: matrix form}
    \end{align}
    where $\bar{\mathsf{x}}^{k+1}$, $\bar{\mathsf{\lambda}}^{k+1}$, and $\bar{u}^{k}$ are augmented vectors defined by $\bar{\mathsf{x}}^{k+1} = [(\mathsf{x}^{k+1})^\top\, \cdots\,(\mathsf{x}^{k+T})^\top]^{\top}$, $\bar{\lambda}^{k+1} = [(\lambda^{k+1})^\top\, \cdots\,(\lambda^{k+T})^\top]^{\top}$ and
    $\bar{u}^{k} = [(u^{k})^\top\, \cdots\,(u^{k+T-1})^\top]^{\top}$.

    It is evident that $E_{12}$ is invertible, as it is an upper triangular matrix with non-zero diagonal entries. Given that the matrix $Q$ is a positive semidefinite, and the matrix $R$ is a positive definite, the matrix $E$ is invertible, and the each block matrix in $E^{-1}$ is obtained by
    \begin{align}
        E^{-1} =  \begin{bmatrix}
        (E^{-1})_{11} & (E^{-1})_{12} & (E^{-1})_{13} \\
        (E^{-1})_{12}^{\top} & (E^{-1})_{22} & (E^{-1})_{23}\\
        (E^{-1})_{13}^{\top} & (E^{-1})_{23}^{\top} & (E^{-1})_{33}
        \end{bmatrix},\label{eqn: inverse of E}
    \end{align}
    where
    \begin{align}
        (E^{-1})_{11}&={(E_{12}^{-1})}^{\top}E_{23}\overline{E}E_{23}^\top E_{12}^{-1},\nonumber
        \\(E^{-1})_{12}&={(E_{12}^{-1})^\top}-{(E_{12}^{-1})^\top}E_{23}\overline{E}E_{23}^{\top}E_{12}^{-1}E_{11}{(E_{12}^{-1})^\top},\nonumber
        \\(E^{-1})_{13}&=-{(E_{12}^{-1})^\top}E_{23}\overline{E},\nonumber
        \\(E^{-1})_{22}&=E_{12}^{-1}E_{11}{(E_{12}^{-1})^\top}(E_{23}\overline{E}E_{23}^\top E_{12}^{-1}E_{11}{(E_{12}^{-1})^\top}-\mathbf{I}),\nonumber
        \\(E^{-1})_{23}&=E_{12}^{-1}E_{11}{(E_{12}^{-1})^\top}E_{23}\overline{E},\nonumber
        \\(E^{-1})_{33}&=\overline{E}\triangleq (E_{33}+E_{23}^\top E_{12}^{-1}E_{11}{(E_{12}^{-1})^\top}E_{23})^{-1}.\nonumber
    \end{align}
    
From the equation \eqref{eqn: matrix form} and \eqref{eqn: inverse of E}, the optimal augmented input vector is determined as
$    \bar{u}^k = -\overline{E}E_{23}^{\top}E_{12}^{-1}F_{1} +\overline{E}E_{23}^{\top}E_{12}^{-1}E_{11}(E_{12}^{-1})^\top F_{2}
$. 
The optimal control input at time $k$ is finally obtained by
$
u^{\star|k} = \begin{bmatrix}
        \mathbf{I}_m & \mathbf{0} & \cdots & \mathbf{0}
        \end{bmatrix}\bar{u}^k
$.
\end{proof}
After the optimal input at time $k$ is applied to the spraying drone, it moves to a new location $y^{k+1}$. To ensure a collaborative behavior of the multi-drone system for smart agriculture, the coverage status must be updated and shared with other drones, as explained in the next section.

\subsection{Stage B. Weight Update Stage}
In D$^2$OC, each drone continuously updates the coverage progress as it sweeps across the farm. This stage is dedicated to maintaining real-time coverage status by dynamically adjusting the weights of nearby sample-points as regions are covered. 

As discussed in Section \ref{sec:2.3}, each agent-point carries a weight $\alpha^k$ at each time step. The objective is to determine how much weight should be transported from each sample-point to a corresponding agent-point while ensuring mass conservation ${\alpha}^k = \sum_{j}\beta_j^k.$

To determine which sample-points to select and how much weight to subtract, the following update equation is applied $\beta^{k+1}_{j} = {\beta}^{k}_{j} - {\gamma}_{j}^{\star\vert k}, \, \forall j,$
where \(\beta^{k}_{j}\) represents the weight of the \(j^{th}\) sample-point, and \(\gamma_{j}^{\star\vert k}\) is the optimal transport plan, which can be efficiently computed by solving the linear programming (LP) problem as described in \cite{kabir2021wildlife} (details omitted for brevity). This update reduces the weights of the sample-points as the agent covers specific areas.

\subsection{Stage C. Weight Sharing Stage}

Since the proposed D$^2$OC framework is designed for decentralized control, the coverage progress of the farm, represented by the weight ${}^r{\beta}^{k}_{j}$ of each sample-point, may differ between agents. Stage C enables the sharing of coverage information between agents by facilitating information exchange among nearby agents within their communication range.

In a decentralized setup, two agents, \( r \) and \( s \), exchange weight information only if they are within the communication range \( d_{\text{comm}} \). The weight-sharing mechanism is defined as  
\[
	{{}^r\beta}^{k}_{j} = {^s\beta}^{k}_{j} = \min\left({^r\beta}^{k}_{j},\,{^s\beta}^{k}_{j}\right), \,\forall j.
\]
In this way, any drones within the communication range can share the coverage progress of the farm area, explicitly represented by the weights of the sample-points.

Thus, the collaborative operation of the multi-drone system is governed by the updated weights of the sample-points, ensuring efficient area coverage. Further details can be found in \cite{kabir2021wildlife}.

These three stages are executed sequentially at each discrete time step. By the end of the mission, the unit mass from the sample-points will be fully transported to the agent-points, while preserving the total mass, thereby ensuring compliance with the mass conservation law as described by the final constraint in \eqref{eqn: W_LP}.

\section{Simulations - Smart Agriculture for Weed Control}

To validate the effectiveness of the D$^2$OC algorithm, we applied it to a weed control problem and compared its performance with other widely used methods: the lawn mower (LM) pattern for uniform area coverage \cite{shahrooz2020agricultural, vazquez2022coverage} and Spectral Multi-scale Coverage (SMC), a state-of-the-art technique for non-uniform area coverage.

\subsection{Simulation Setup}

\begin{figure}[!h]
    \centering
    \subfloat[Agricultural spray drone]{\includegraphics[scale=0.28]{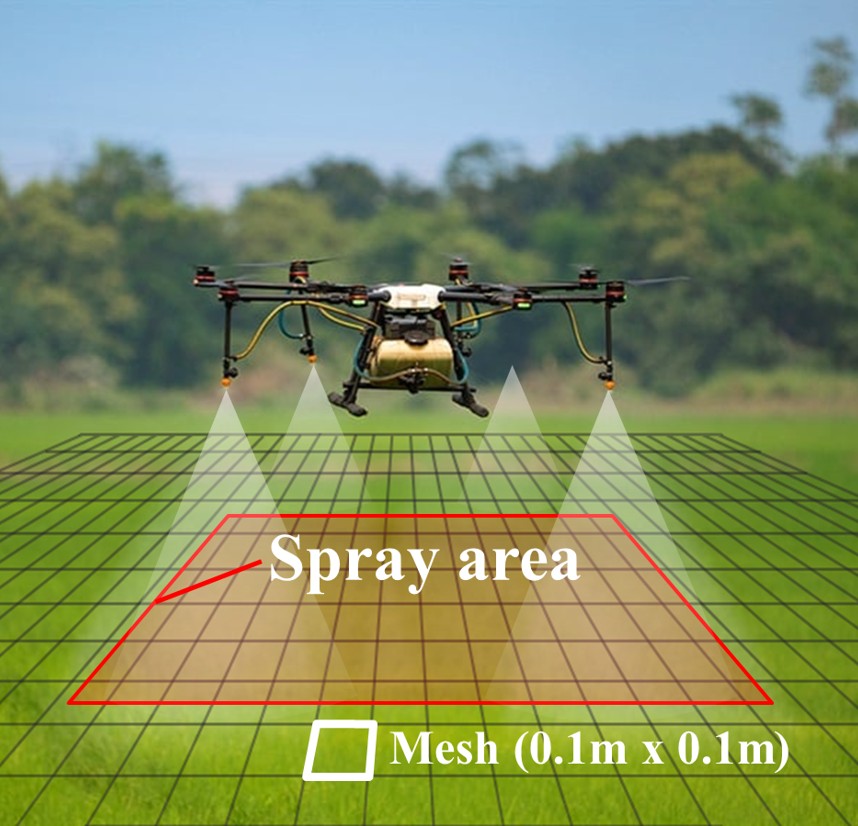}}\\
    \subfloat[Weed density map]{\includegraphics[scale=0.30]{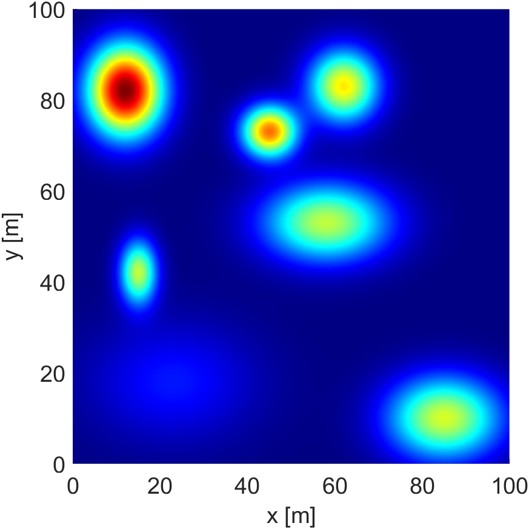}}\quad
    \subfloat[Sample-point distribution]{\includegraphics[scale=0.30]{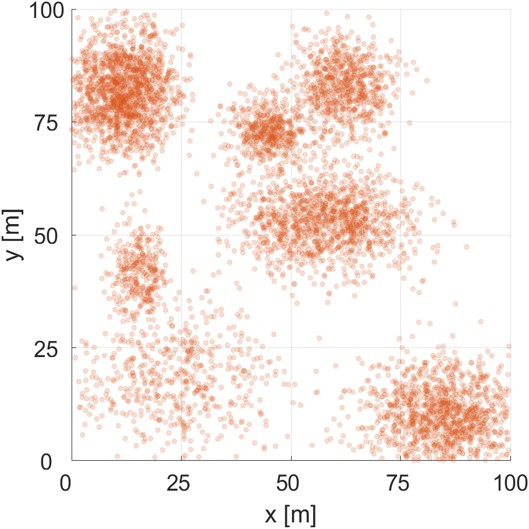}}
    \caption{(a) An agricultural spray drone; 
    (b) Weed density map modeled by a mixture of Gaussian distributions; (c) Sample-point distribution obtained from (b).}
    \label{fig: agri_drone + ref. PDF}
\end{figure}

The simulation setup involves agricultural drones, as shown in Fig. \ref{fig: agri_drone + ref. PDF}(a), deployed for weed management across a farm. These drones are guided by a weed map obtained using either satellite or UAV imagery, which is assumed to be available a priori. The weed density map is modeled as a Gaussian mixture in \ref{fig: agri_drone + ref. PDF}(b), and the corresponding sample-point distribution is obtained through sampling techniques, as shown in \ref{fig: agri_drone + ref. PDF}(c).

In real-world farming environments, weeds coexist with crops, making their management crucial for controlling spread while minimizing herbicide use. Over-application of chemicals can damage crops and harm the environment, which underscores the need for more targeted spraying solutions. Traditional spraying methods often apply chemicals uniformly across both weed-infested and healthy areas, leading to inefficiencies. This highlights the need for a more intelligent, multi-agent control scheme, such as D$^2$OC, to optimize chemical usage and enhance weed suppression effectiveness.

The simulations utilize the DJI AGRAS T10 agricultural spraying drone over a $100\ \text{m} \times 100\ \text{m}$ farm. Although this farm size is relatively small, the operation time is deliberately constrained to 180 and 300 s to assess the scalability of our method and its applicability to larger-scale UAV operations.  

A key strength of the D$^2$OC framework is its ability to scale efficiently. By leveraging a reference density map that adapts to different farm sizes, the method remains inherently \textit{scale-independent}. This ensures that larger farms can utilize proportionally scaled reference densities and drone task distributions without altering the underlying methodology. Moreover, its modular design allows seamless expansion to larger agricultural operations by adjusting the number of UAVs or flight durations.

The domain is discretized into grid cells of size $0.1~\text{m} \times 0.1~\text{m}$. The weed density at each cell is assigned according to the density function shown in Fig. 6(b) and normalized such that the maximum density among all cells is 1, which occurs at $(x, y) = (12~\text{m}, 82~\text{m})$.
The initial and final (or survived) weed densities of grid cell $c$ are denoted by $\rho_{c}^{(0)}$ and $\rho_{c}^{(f)}$, respectively. This setup enables tracking of herbicide dose and weed survival for each grid cell.

Fig. \ref{fig: agri_drone + ref. PDF}(a) illustrates the spraying model adopted in the simulations. The herbicide was assumed to be uniformly distributed over a square spray area, with the side length determined by the flight altitude, as summarized in Table~\ref{table: parameter}. Environmental factors such as wind drift and spray system characteristics, which can significantly affect spray dispersion in practice, were not incorporated. This simplification was made to isolate and evaluate the 
collaborative behavior and coverage performance of the proposed controller under controlled conditions. The total herbicide dose in each grid cell $c$, denoted by $x_c$, was then computed as the cumulative sum of all doses applied over the course of the simulation.

To further enhance realism, agricultural parameters were drawn from \cite{ritz2015research}, which includes weed and herbicide types, as well as the relationship between herbicide concentration and weed survival. Biotype 289 is selected as the weed type, and glyphosate is used as the herbicide. The weed survival in response to herbicide dose follows a sigmoid function, $
\rho_{c}^{(f)} = \dfrac{\rho_{c}^{(0)}}{1 + \exp\big(\log(x_c) + \log(LD_{50})\big)}$,
where \(LD_{50}\) is the survival rate parameter (Table \ref{table: parameter}), and $x_c$ is the total applied herbicide dose in grid cell $c$.

\begin{table}[!h]
\centering
\renewcommand{\arraystretch}{0.9} 
\setlength{\tabcolsep}{4pt} 
\begin{tabular}{llcl}
\hline
Category & Parameter & Symbol & Value \\ \hline
\multirow{7}{*}{Drone}  
    & Number of drones & $n_d$ & 3  \\  
    & Mass & $m_d$ & 0.3 kg \\  
    & Inertia (x, y) & $I_{d,x}, I_{d,y}$ & 0.2 kg$\cdot$m$^2$ \\  
    & Inertia (z) & $I_{d,z}$ & 0.4 kg$\cdot$m$^2$ \\  
    & Max thrust & $f_{th,m}$ & 440 N \\  
    & Max torque (roll/pitch) & $\tau_{\phi,m}, \tau_{\theta,m}$ & 81.4 N$\cdot$m \\  
    & Max torque (yaw) & $\tau_{\psi,m}$ & 5.5 N$\cdot$m \\  
    & Max speed & $v_m$ & 7 m/s \\  
    & Max angle (roll/pitch) & $\phi_m, \theta_m$ & 15 ° \\  
    & Max angular vel. & $|\omega|_{max}$ & 15 °/s \\  
    & Operation time & & 180 s, \, 300 s \\ \hline
\multirow{3}{*}{Spraying}  
    & Tank dimensions & $l_L, l_D$ & 0.15, 0.2 m \\  
    & Spray rate & $Q_s$ & 1.4$\times 10^{-5}$ m$^3$/s \\  
    & Spray width (altitude) & & 3–5.5 m (1.5–3 m) \\ \hline
\multirow{5}{*}{\shortstack[l]{Weed \& \\ Herbicide}}  
    & Weed type & & biotype 289 \\  
    & Herbicide & & glyphosate \\  
    & Survival rate & $LD_{50}$ & 134.2 \\  
    & Concentration (0.7\%) & & 495.3 g/m$^3$ \\  
    & Solution density & $\rho_s$ & 1000 kg/m$^3$ \\ \hline
    \multirow{4}{*}{D$^2$OC}  
    & Horizon length & $T$ & 60 \\  
    & Penalty matrices &\multicolumn{2}{l}{$\begin{array}{l}
      Q = 10^{-7} \times \operatorname{diag}(1,1,1,1,1,1,\\
      \qquad\ 10^3,10^3,10^3,0,0,10^3) \\
      R = 10^{-3} \times \operatorname{diag}(1,1,1,1)
    \end{array}$}\\ \hline
\end{tabular}
\caption{Simulation parameters. All values are in SI units unless otherwise noted.}
\label{table: parameter}
\end{table}

We tested three different spraying methods: Lawn Mower (LM), Spectral Multi-scale Coverage (SMC), and D$^2$OC. While LM is a commonly used method for uniform coverage \cite{shahrooz2020agricultural, vazquez2022coverage}, both SMC and D$^2$OC are designed to address non-uniform area coverage challenges.

\subsection{Performance Comparison between the LM, SMC, and D$^2$OC Methods}
A total of three drones were considered throughout the simulations. For the LM method, the farm areas need to be divided into three regions.
Although the weed reference map is available, the LM method cannot incorporate this information into the plan because it was mainly for uniform area coverage. Thus, the region was divided into three equal areas, each of which was assigned to a drone.
For each subregion, the reference trajectory was generated for each drone. To ensure uniform coverage and timely completion, the number of waypoints was determined based on the operation time, and the waypoints were evenly spaced along the planned path. A model predictive control (MPC) was employed to track the reference trajectory.

The SMC method was implemented in the simulation with a two-stage cascaded control structure: at each time step, the reference path over a specified horizon was obtained using the SMC method under first-order dynamics, and an MPC controller was then employed to track it. It is worth noting that the control law for the LTV model cannot be directly obtained from the SMC method since it was developed specifically for first- and second-order integrators. Therefore, in the simulations, the MPC controller served as the low-level tracking controller. The SMC method employed 40 Fourier cosine bases for each of the $x$- and $y$-axes. The MPC horizon length was set to 20.

In contrast to the SMC method, the D$^2$OC method was employed in the simulation directly under a linearized drone model. The parameters used for the D$^2$OC are shown in Table \ref{table: parameter}. Importantly, the 10th and 11th diagonal elements of the matrix $Q$, corresponding to $x$- and $y$-coordinates, are set to zero. By setting zero penalty on the agent’s 
$x$- and $y$-coordinates in the cost function, the controller focuses on other objectives, without directly regulating its position. To assess the performance of D$^2$OC schemes with and without a communication range constraint, simulations were conducted under two configurations: centralized communication and a 10 m communication range.

\begin{figure*}[!h]
	\centering
	\subfloat[LM]{\includegraphics[width=0.230\linewidth]{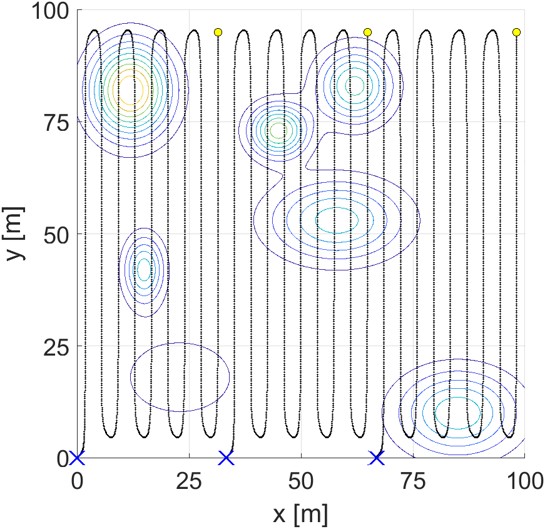}}\quad
	\subfloat[SMC]{\includegraphics[width=0.230\linewidth]{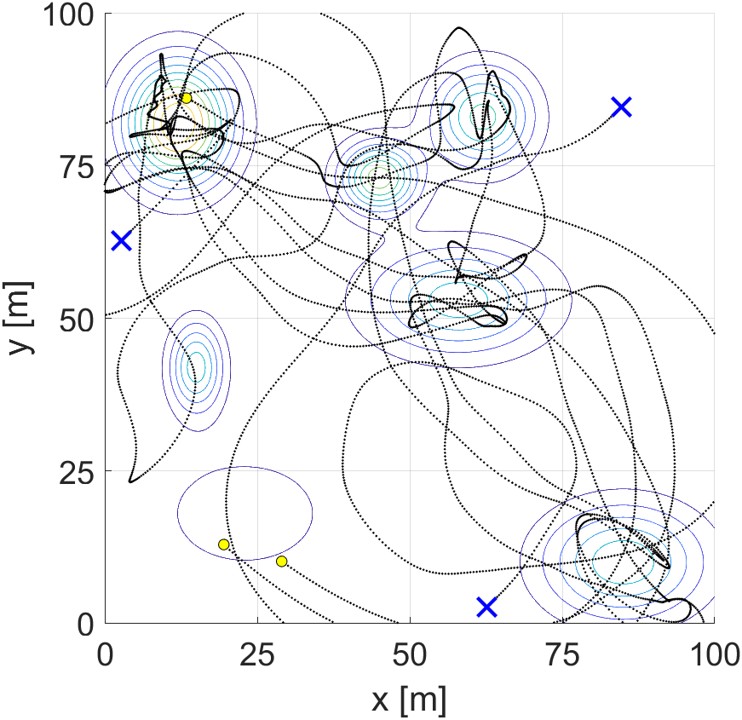}}\quad
	\subfloat[D$^2$OC, Centralized communication]{\includegraphics[width=0.230\linewidth]{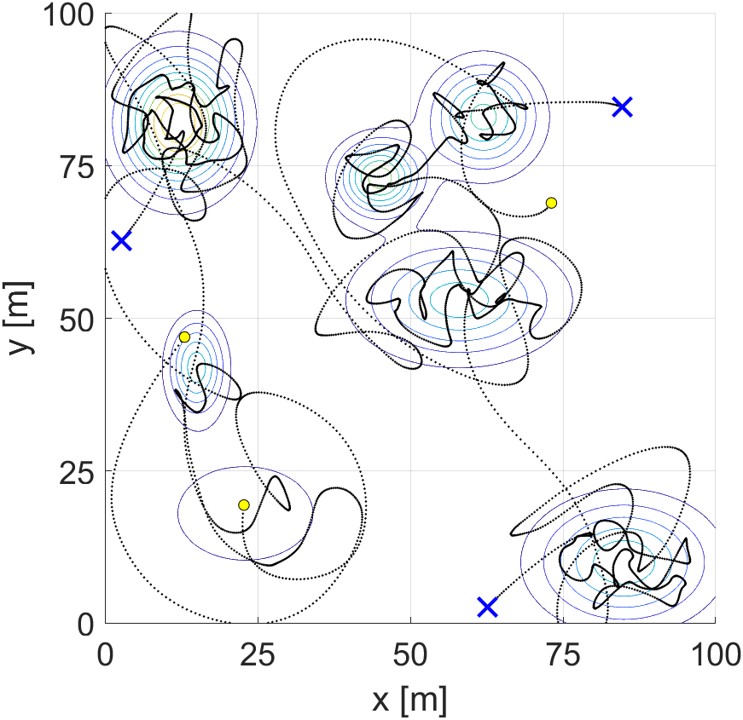}}\quad
    \subfloat[D$^2$OC, Decentralized with $r_\text{comm}=10 \text{ m}$]{\includegraphics[width=0.230\linewidth]{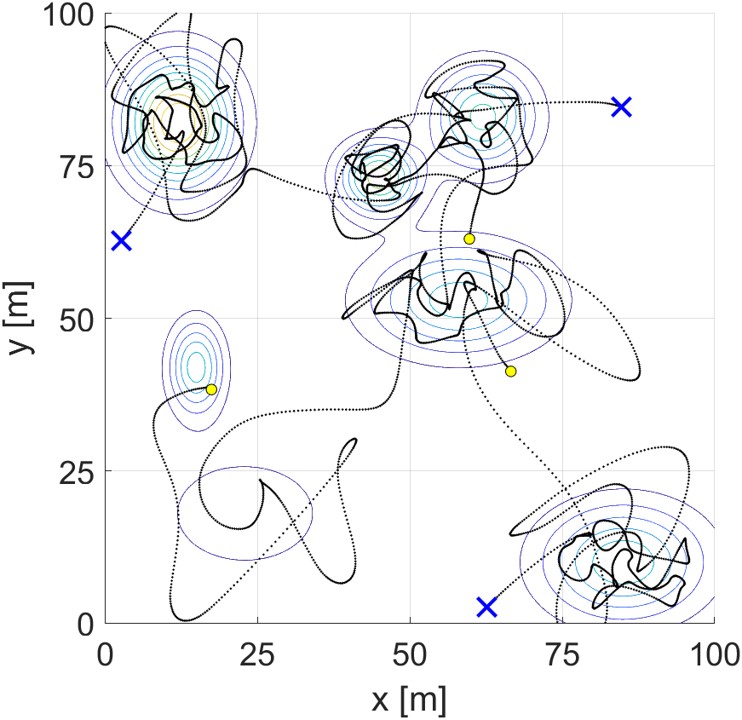}}
    \caption{Trajectories for four methods/settings (operation time: 180 s).}\label{fig: comparison_Traj_180}
    \end{figure*}
    \begin{figure*}[!h]
	\subfloat[LM]{\includegraphics[width=0.230\linewidth]{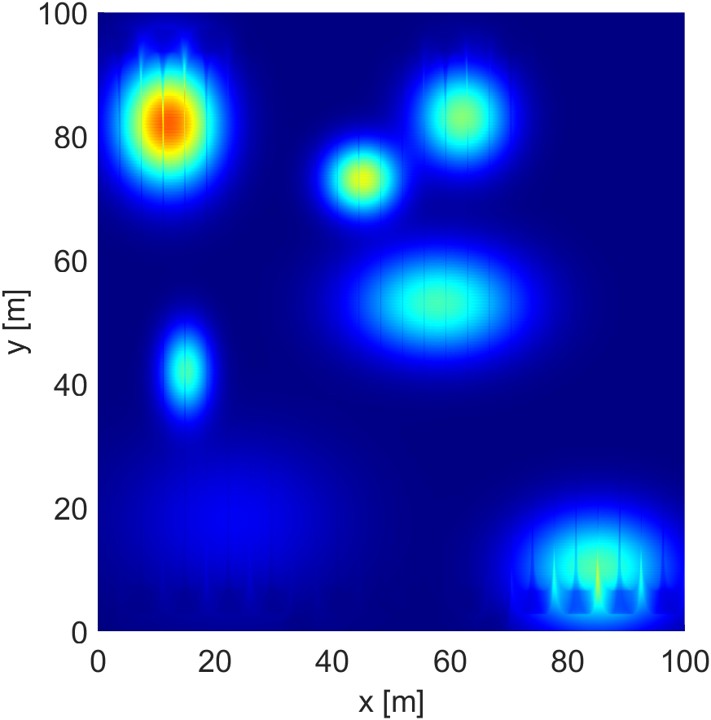}}\quad
	\subfloat[SMC]{\includegraphics[width=0.230\linewidth]{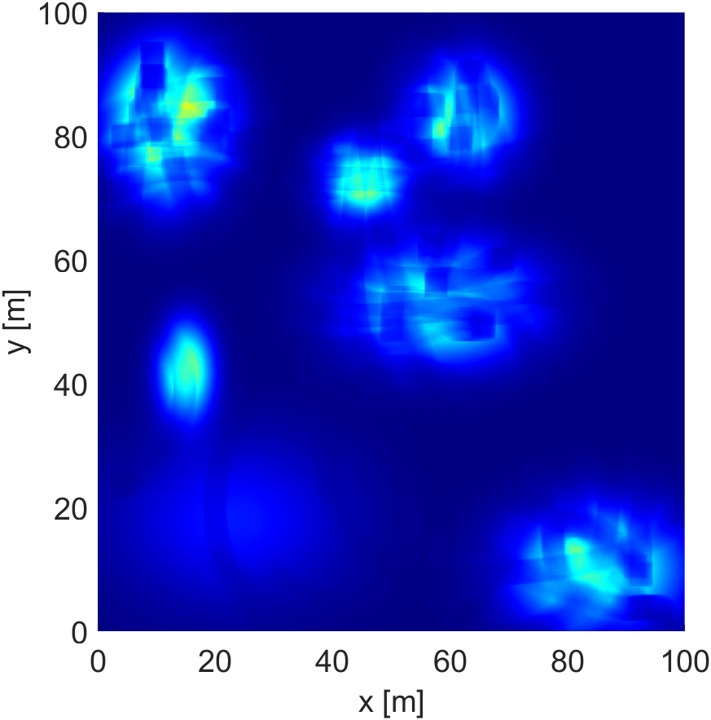}}\quad
	\subfloat[D$^2$OC, Centralized communication]{\includegraphics[width=0.230\linewidth]{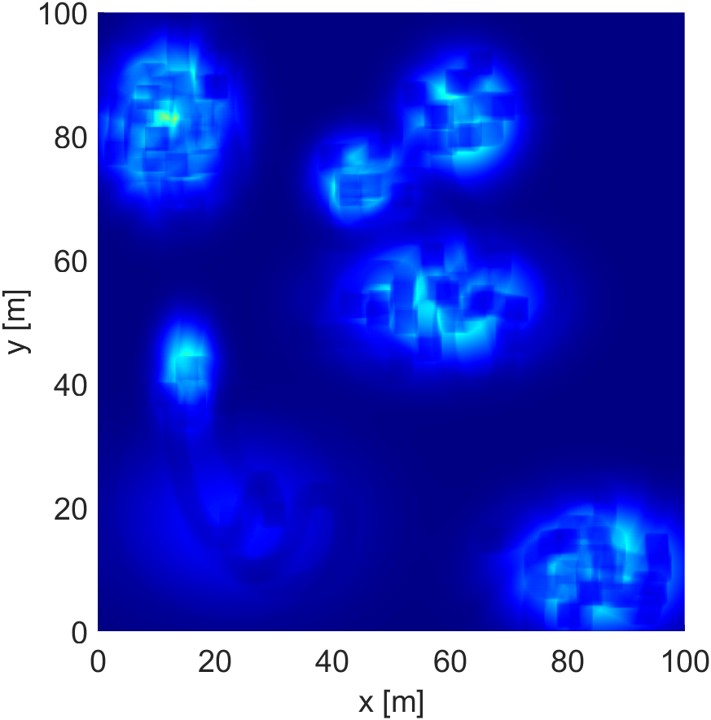}}\quad
    \subfloat[D$^2$OC, Decentralized with $r_\text{comm}=10 \text{ m}$]{\includegraphics[width=0.230\linewidth]{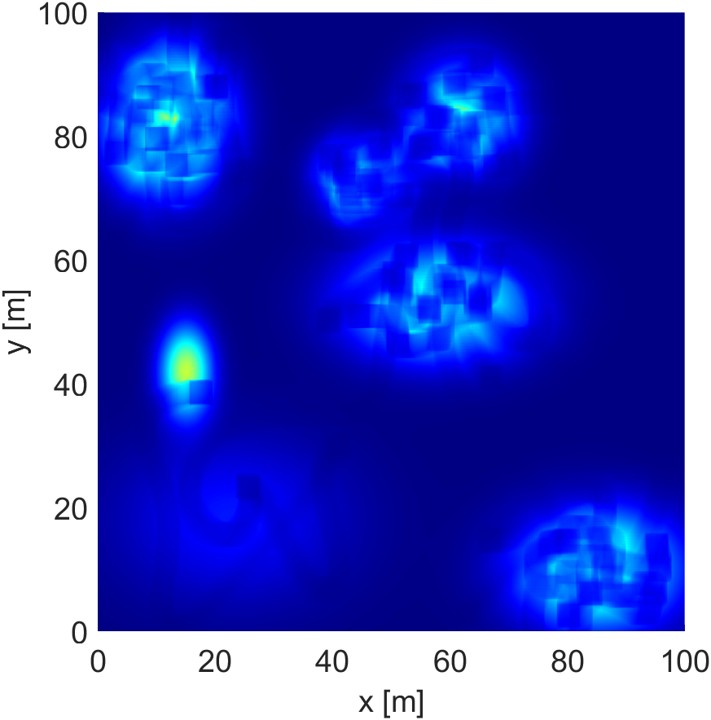}}\\
	\caption{Colormap of weed survival density for four methods/settings (operation time: 180 s, red: highest, blue: lowest).}\label{fig: comparison_Weed_180}
\end{figure*}

\begin{figure*}[!t]
\vspace{0.01in}
    \centering
	\subfloat[LM]{\includegraphics[width=0.230\linewidth]{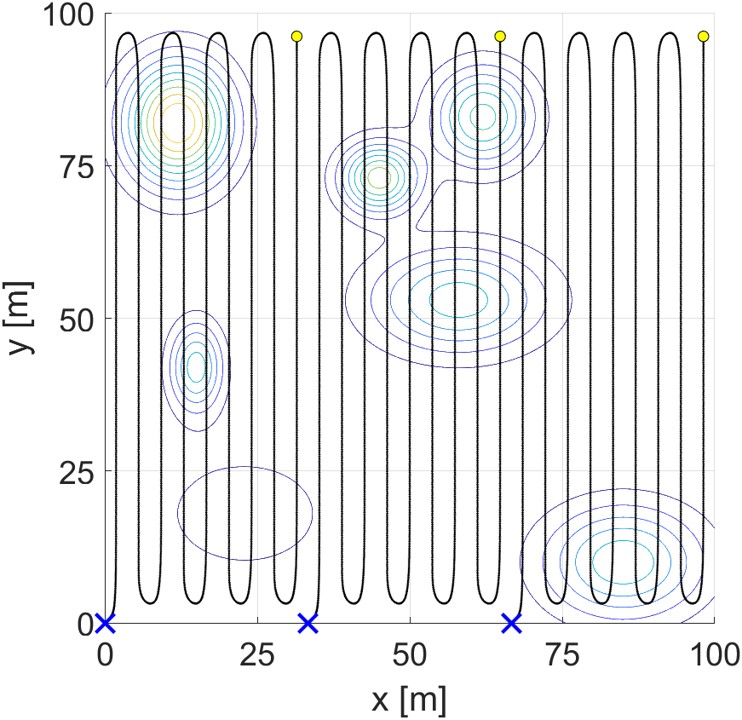}}\quad
	\subfloat[SMC]{\includegraphics[width=0.230\linewidth]{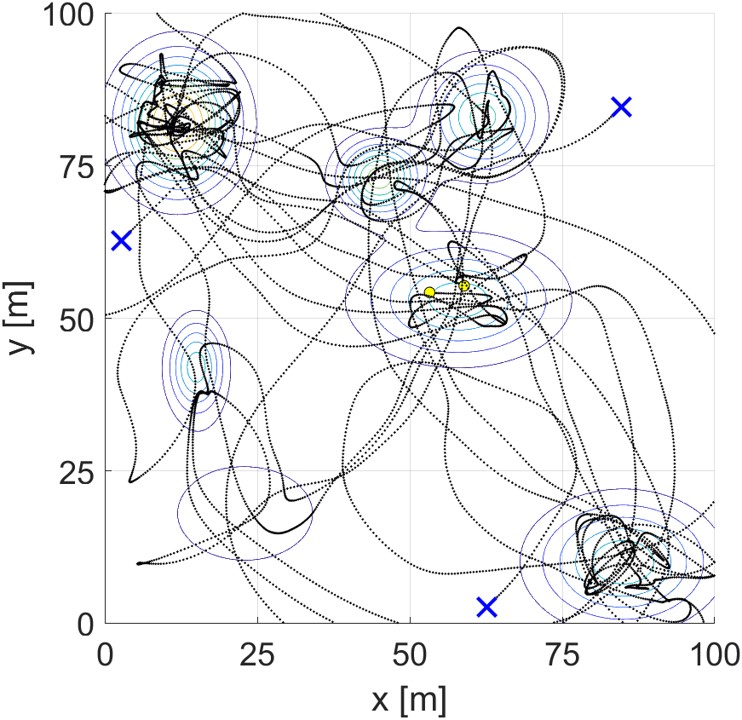}}\quad
	\subfloat[D$^2$OC, Centralized communication]{\includegraphics[width=0.230\linewidth]{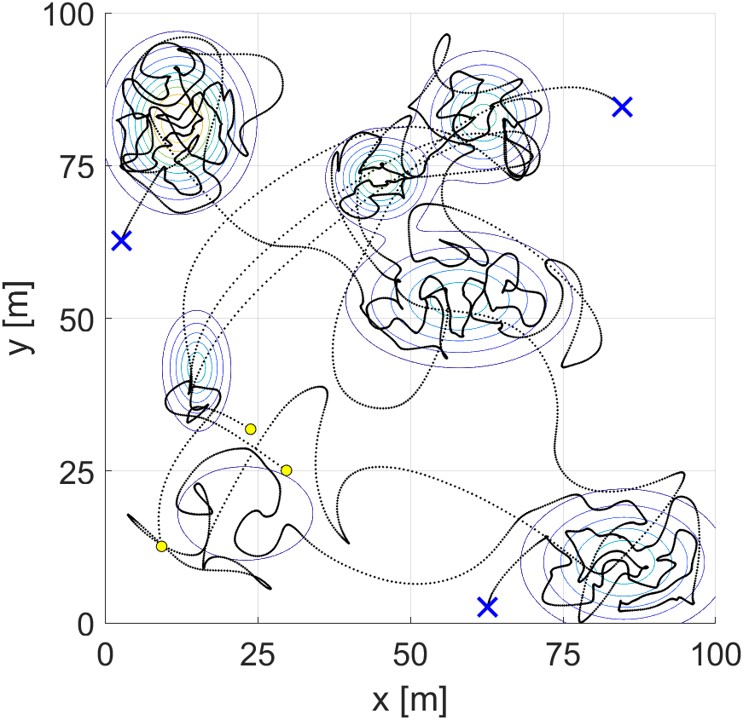}}\quad
	\subfloat[D$^2$OC, Decentralized with $r_{\text{comm}}=10\text{ m}$]{\includegraphics[width=0.230\linewidth]{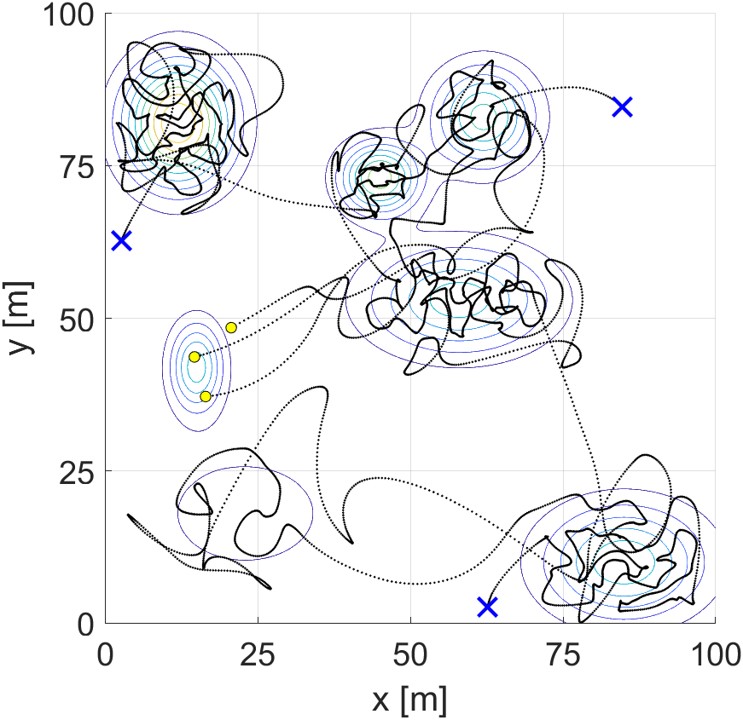}}
    \caption{Trajectories for four methods/settings (operation time: 300 s).}\label{fig: comparison_Traj_300}
\end{figure*}
\begin{figure*}[!t]
	\subfloat[LM]{\includegraphics[width=0.230\linewidth]{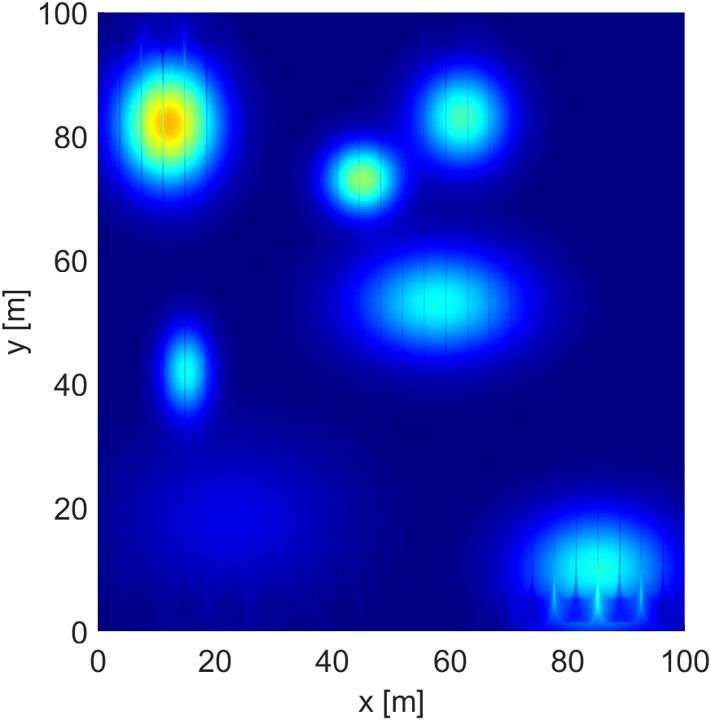}}\quad
	\subfloat[SMC]{\includegraphics[width=0.230\linewidth]{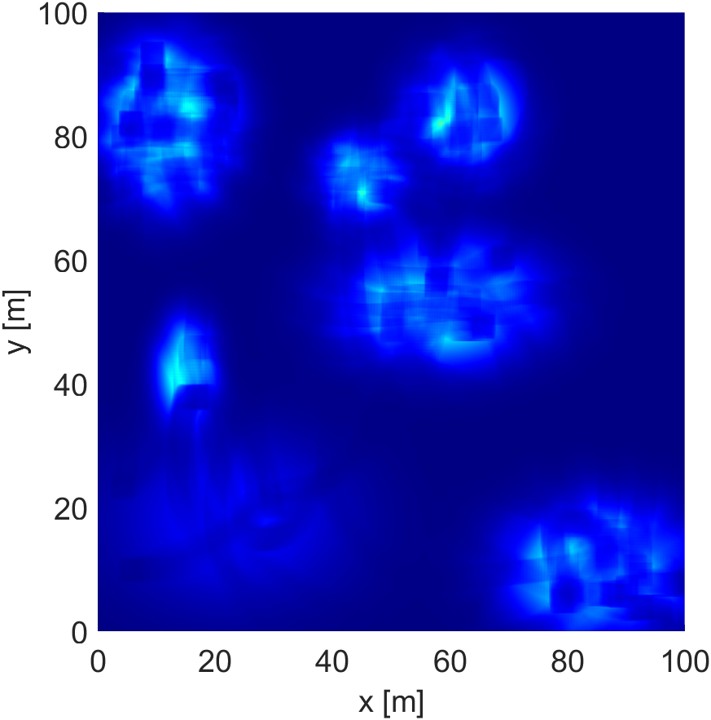}}\quad
	\subfloat[D$^2$OC, Centralized communication]{\includegraphics[width=0.230\linewidth]{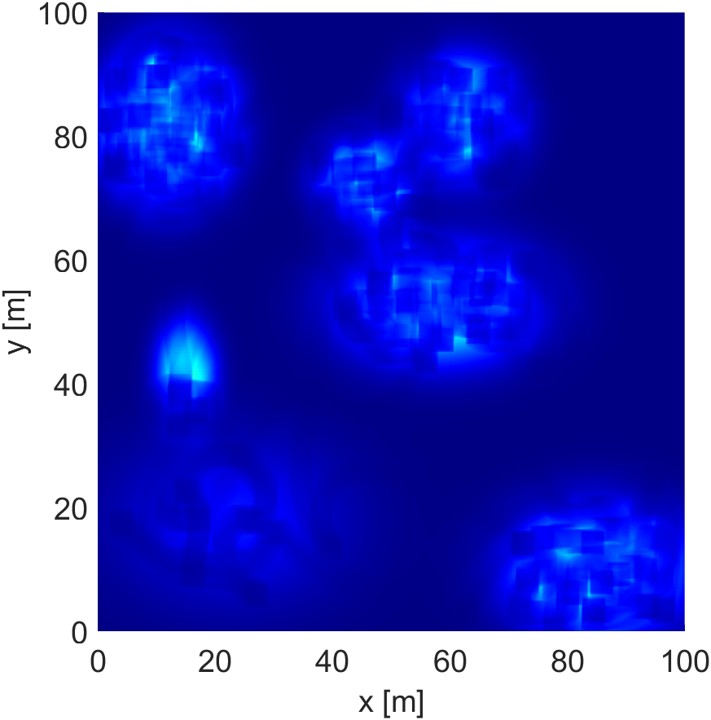}}\quad
	\subfloat[D$^2$OC, Decentralized with $r_\text{comm}=10\text{ m}$]{\includegraphics[width=0.230\linewidth]{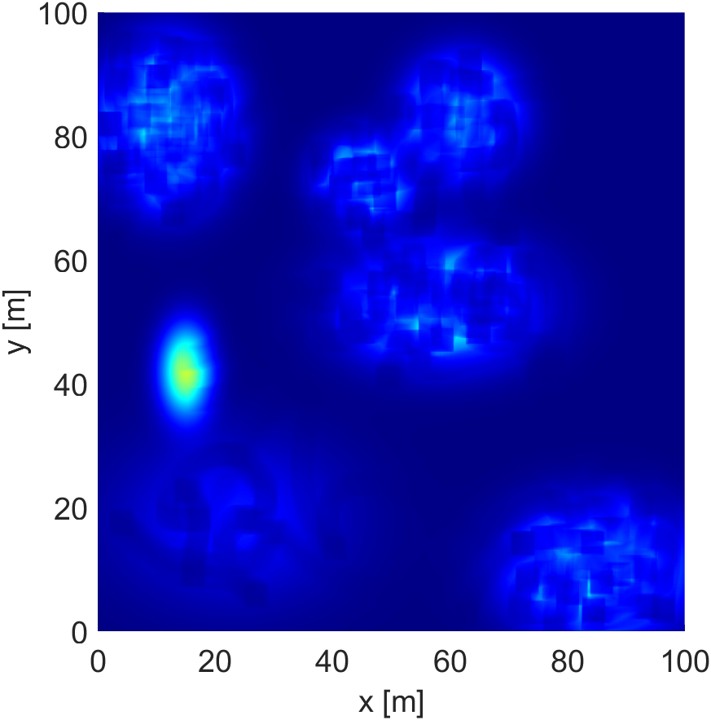}}
	\caption{Colormap of weed survival density for four methods/settings (operation time: 300 s, red: highest, blue: lowest).}\label{fig: comparison_Weed_300}
\end{figure*}

The simulations were conducted for two different operation times: 180 and 300 s. 
The simulation with a longer operation time is expected to kill more weeds, as drones will spray more herbicide at the same rate over a longer period. Figs. \ref{fig: comparison_Traj_180}, \ref{fig: comparison_Weed_180}, \ref{fig: comparison_Traj_300}, and \ref{fig: comparison_Weed_300} provide the graphical performance comparison for the operation times of 180 and 300 s. 
Figs. \ref{fig: comparison_Traj_180} and \ref{fig: comparison_Traj_300} present the resulting trajectories of the agents. 
The blue cross symbols represent the initial position of three drones, and the yellow circle symbols indicate their final positions. Figs. \ref{fig: comparison_Weed_180} and \ref{fig: comparison_Weed_300} show the survived weed density after the herbicide is applied, where the initial weeds are shown in Fig. \ref{fig: agri_drone + ref. PDF}(b).

The LM method, shown in Figs.~\ref{fig: comparison_Traj_180}(a) and \ref{fig: comparison_Traj_300}(a), covers the entire area uniformly without considering weed distribution, which leads to inefficient herbicide use, with excessive dosage in low-density areas and insufficient dosage in high-density areas.
Applying excessive herbicides in areas with low weed density tends to harm the environment, including animals, microbes, and soil, while applying insufficient herbicides in areas with high weed density is ineffective for controlling the weeds. 

On the contrary, the agent's trajectories using the SMC and D$^2$OC methods demonstrate that the agents spent more time flying over areas with higher weed density.
Compared to the trajectories of the D$^2$OC method, those of the SMC method showed the cursory coverage of the weed-concentrated areas. This can be attributed to several possible factors. 
Firstly, the SMC method under first-order dynamics was used as a high-level controller to generate the reference path, and an MPC was employed to track it. As a result, the agent may deviate from the reference path, which can degrade coverage performance.
Secondly, the SMC method does not consider the available operation time in obtaining the control law. In the SMC method, the control law is designed as the gradient that minimizes the ergodicity between agents' trajectories and the given spatial distribution, while disregarding the available operation time. The comparison of Fig. \ref{fig: comparison_Traj_180}(b) with Fig. \ref{fig: comparison_Traj_300}(b) effectively shows this limitation. The trajectories in Fig. \ref{fig: comparison_Traj_180}(b), corresponding to an operation time of 180 s, coincide with the initial segment of those in Fig. \ref{fig: comparison_Traj_300}(b), corresponding to an operation time of 300 s, indicating the method's independence of operation time.
The operation time-independent ergodic nature of the SMC method caused the agent to frequently traverse low-priority areas between high-density regions, leading to resource wastage and unintended chemical spraying.

On the other hand, the trajectory of the D$^2$OC in Fig. \ref{fig: comparison_Weed_180}(c) does not coincide with that in Fig. \ref{fig: comparison_Weed_300}(c), demonstrating the method's consideration of operation time. The D$^2$OC framework allows agents to allocate their effort according to the operation time, achieving more efficient coverage within the given operation time. This inherent feature led to trajectories that better aligned with the weed density distribution, resulting in fewer surviving weeds compared to the LM and SMC methods, as shown in Figs. \ref{fig: comparison_Weed_180}(c) and \ref{fig: comparison_Weed_300}(c), indicating a higher weed-killing rate.

Figs. \ref{fig: comparison_Traj_180}(d) and \ref{fig: comparison_Traj_300}(d) show the agent trajectories for the D$^2$OC method with a limited communication range. The trajectories are largely similar to those under centralized communication. However, differences appear toward the end of the trajectories. Due to the limited communication between agents, they were unable to fully share and synchronize their weight information, leading to a lack of awareness of others' coverage. Consequently, some agents allocated coverage effort to areas that had already been covered by others, rather than to the Gaussian distribution centered at $x=20$ and $y=40$.

To quantitatively evaluate the performances of the methods, two metrics (the maximum survival density and the reduction rate) are defined, and the results are presented in Tables \ref{table: comparison_180} and \ref{table: comparison_300}.
\begin{table}
\centering
\begin{minipage}{1\linewidth}
\centering
\begin{tabular}{lccc}
\hline
\textbf{Method} & \begin{tabular}[c]{@{}c@{}}Total Herbicide\\ Dosage (g ai)\end{tabular} &
\begin{tabular}[c]{@{}c@{}}Max. Survival\\Density\end{tabular} &
\begin{tabular}[c]{@{}c@{}}Reduction\\ Rate (\%)\end{tabular} \\
\hline\hline
LM     & 38.12 & 0.795 & 22.07 \\
SMC    & 37.13 & 0.638 & 31.92 \\
D$^2$OC (centralized) &37.13 & 0.573 & 38.04 \\
D$^2$OC (decentralized) & 37.13 & 0.573 & 37.87 \\
\hline
\end{tabular}
\caption{Performance comparison of the methods after 180 s of operation.}
\label{table: comparison_180}
\end{minipage}

\begin{minipage}{1\linewidth}
\vspace{0.1in}
\centering
\begin{tabular}{lccc}
\hline
\textbf{Method} & \begin{tabular}[c]{@{}c@{}}Total Herbicide\\ Dosage (g ai)\end{tabular} &
\begin{tabular}[c]{@{}c@{}}Max. Survival\\Density\end{tabular} &
\begin{tabular}[c]{@{}c@{}}Reduction\\ Rate (\%)\end{tabular} \\
\hline\hline
LM     & 62.87 & 0.694 & 31.76 \\
SMC    & 62.04 & 0.479 & 44.90 \\
D$^2$OC (centralized) & 62.04 & 0.407 & 50.18 \\
D$^2$OC (decentralized) & 62.04 & 0.563 & 49.28 \\
\hline
\end{tabular}
\caption{Performance comparison of the methods after 300 s of operation.}\label{table: comparison_300}
\end{minipage}

\begin{minipage}{1\linewidth}
\vspace{0.1in}
\centering
\includegraphics[width=1\linewidth]{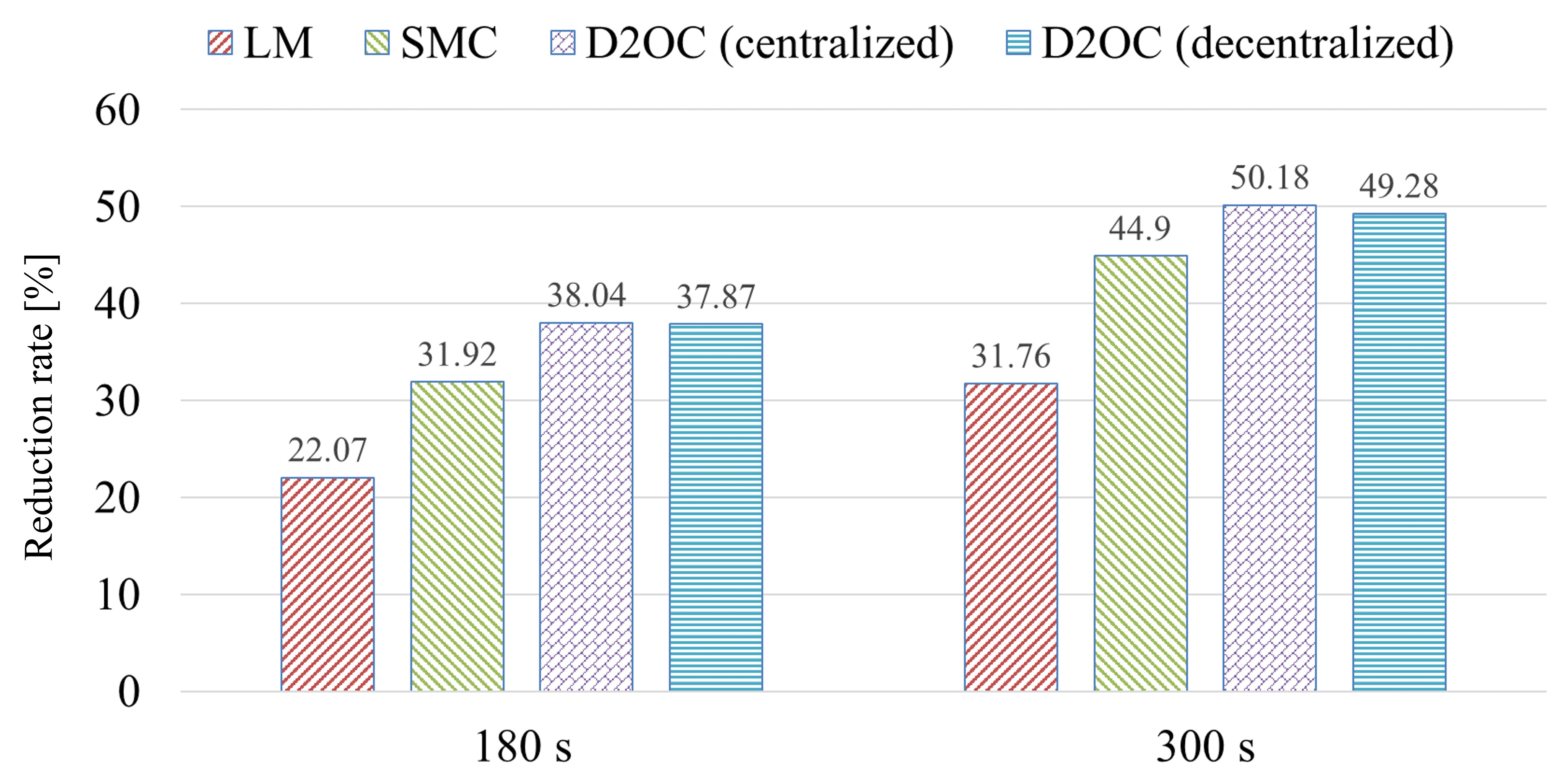}
    \captionof{figure}{Comparison of reduction rates of the methods.}
    \label{fig:enter-label}

    \end{minipage}
\end{table}
The metric \textit{maximum survival density} represents the highest surviving weed density among all grid cells, defined as
\begin{equation*}
    \text{Max. survival density}= \max_{c} \rho^{(f)}_c.
\end{equation*}

The metric \textit{reduction rate} is defined as
\begin{equation*}\label{eqn: reduction rate}
\begin{aligned}
	 \text{Reduction rate}\,(\%) = \frac{\sum\limits_{c}{(\rho^{(0)}_{c}-\rho^{(f)}_{c})}}{\sum\limits_{c}\rho^{(0)}_{c}} \times 100,
\end{aligned}
\end{equation*}denoting how much weed is reduced compared to the initial amount of weed across the entire domain. For instance, the reduction rate of 0\% means that no weed is removed in any area, whereas 100\% means all the weeds are removed in the entire domain. Alternatively, this indicates overall weed removal performance in the given domain.

For all presented performance indices, the D$^2$OC method outperformed other methods, indicating that it covered the non-uniform weed distribution better. 
Furthermore, the difference in reduction rate between the SMC and D$^2$OC methods was 6.12 percentage points at 180 s, and 5.28 percentage points at 300 s, as highlighted in Fig. \ref{fig:enter-label}. 
The D$^2$OC under limited communication has a slightly lower reduction rate compared to the D$^2$OC under centralized communication, although it is still higher than that of the SMC method. Given that the SMC method considered centralized communication, this result effectively shows the practicality of the D$^2$OC method in real-world applications.

As the operation time increases indefinitely, the reduction rate is anticipated to approach 100\%. However, for finite operation times, the D$^2$OC method demonstrates a higher reduction rate compared to the SMC method. This advantage arises from D$^2$OC's incorporation of operation time into its control strategy. The flexibility of D$^2$OC in managing both energy consumption and operation time makes it well-suited for real-world smart agricultural applications, where drones or agents are constrained by limited energy and operational time, and the farm size is not negligible.

\begin{figure}[!h]
    \centering
    \subfloat[Initial weed]{\includegraphics[width=0.5\linewidth]{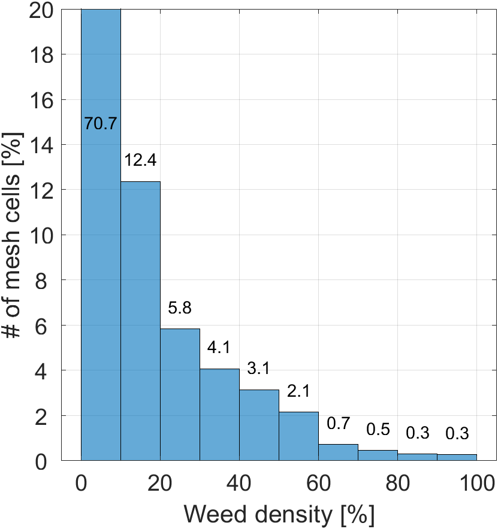}}
    \subfloat[LM]{\includegraphics[width=0.5\linewidth]{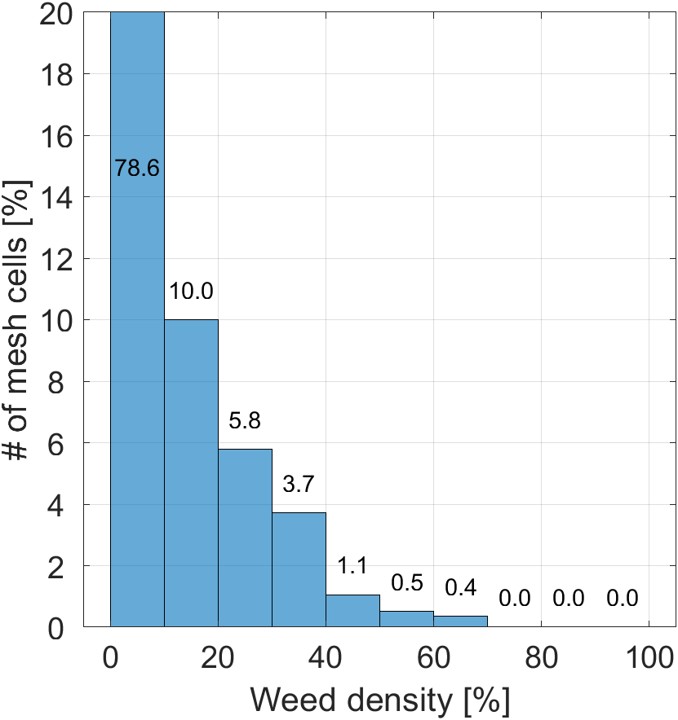}}
    \\
    \subfloat[SMC]{\includegraphics[width=0.5\linewidth]{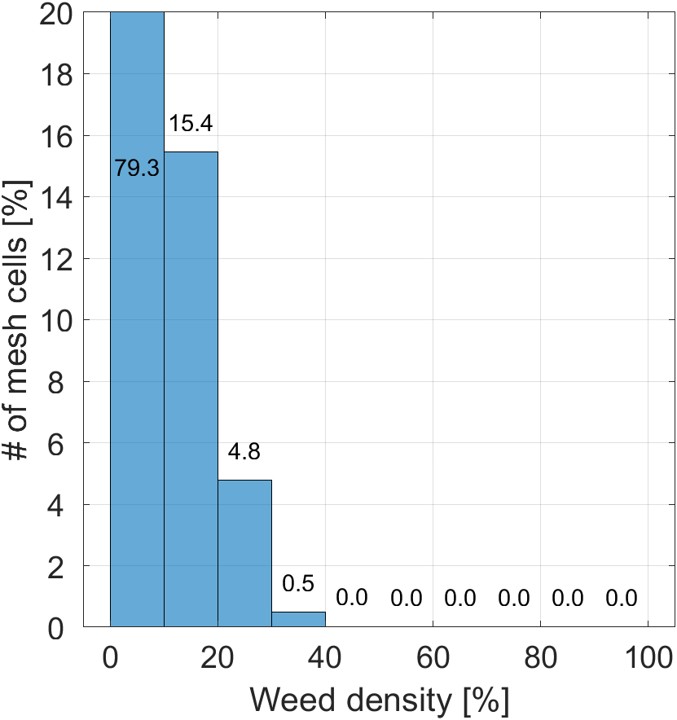}}
    \subfloat[D$^2$OC, centralized communication]{\includegraphics[width=0.5\linewidth]{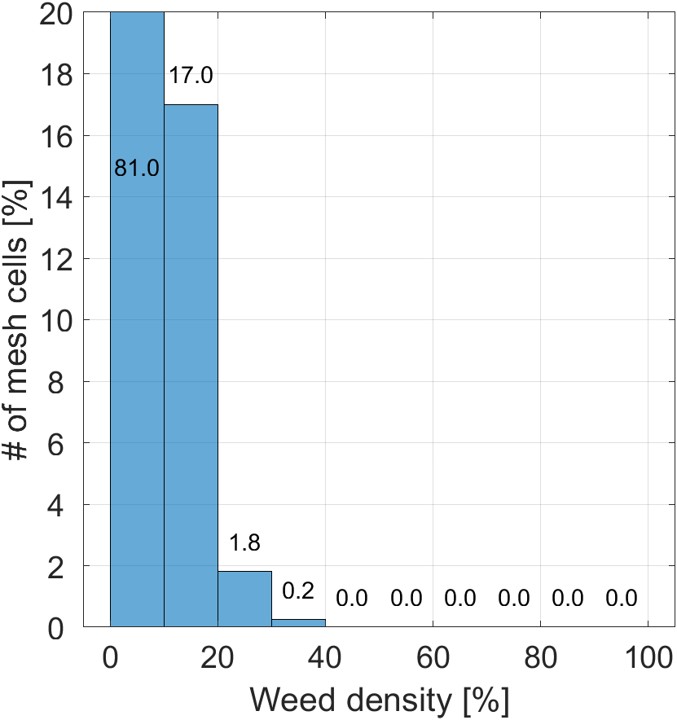}}
    \caption{Histograms of the weed survival when the operation time is 300 s.}
    \label{fig: hist}
\end{figure}
Fig. \ref{fig: hist} presents the histograms of the weed survival when the operation time is 300 s. The number of grid cells is displayed as a percentage, with weed density intervals increasing by 10 percent. Fig. \ref{fig: hist}(a) shows the histogram of the initial weed distribution. Figs. \ref{fig: hist}(b, c, d) illustrate the histogram of the weed survival after the chemical-spraying task using the LM, SMC, and D$^2$OC methods, respectively. 
The LM method shows a relatively consistent reduction in weed density across the histogram bins in Fig. \ref{fig: hist}(b).
However, compared to other methods, more grid cells still have high weed density (greater than 20\%). This outcome is expected since the LM method is based on a uniform area coverage approach and does not take into account the density of the reference. 
The SMC method in Fig. \ref{fig: hist}(c) shows improved weed control relative to the LM method by selectively reducing the weed density in grid cells with higher infestation, demonstrating its non-uniform coverage strategy.
Among the three methods, the D$^2$OC method demonstrates the best non-uniform coverage. Only 2.0 percent of the grid cells exhibit a weed density greater than 20\%. 

Weeds in areas with high weed density have a higher chance of growing larger. Weeds at later growth stages can negatively affect weed control in two ways. Firstly, the later the weed's growth stage, the higher the dosage of herbicide required to kill it \cite{kieloch2011role}. In other words, once the weeds have grown, they become more difficult to control. Furthermore, larger weeds produce more seeds, leading to the reproduction and spread of more weeds across the farm area. From this perspective, the D$^2$OC method showed promising performance in weed management among the methods by flattening the weed distribution density better than other methods.

\section{Conclusion}  
This paper presented the Density-Driven Optimal Control (D$^2$OC) strategy for multi-drone systems in smart agriculture, focusing on scalable pest, weed, and disease management. Spraying drone dynamics were modeled as a Linear Time-Varying (LTV) system to capture mass and inertia variations due to chemical dispersion. The optimal control input was derived using Lagrangian mechanics, with the optimization problem formulated via Wasserstein distance to quantify discrepancies between the reference density map and drone coverage.  

Simulations showed that D$^2$OC enhances farm coverage efficiency over conventional methods. By prioritizing high-risk areas and minimizing unnecessary chemical use, it optimally distributes spraying tasks, reducing resource consumption and improving sustainability. These findings underscore D$^2$OC's potential for large-scale smart farming, boosting productivity while mitigating environmental impact. Future research will extend the D$^2$OC framework to nonlinear UAV dynamics and advanced optimal control strategies, while explicitly considering environmental disturbances and uncertainties in operational conditions, including variations in power availability and localization errors. 
These efforts aim to strengthen the theoretical foundations and enhance performance and robustness in highly dynamic and uncertain scenarios.

\section*{Acknowledgment}

This work was supported by the NSF CAREER Grant CMMI-DCSD-2145810.

\bibliographystyle{ieeetr}
{
\bibliography{references}
}

%

\begin{IEEEbiography}[{\includegraphics[width=1in,height=1.25in,clip,keepaspectratio]{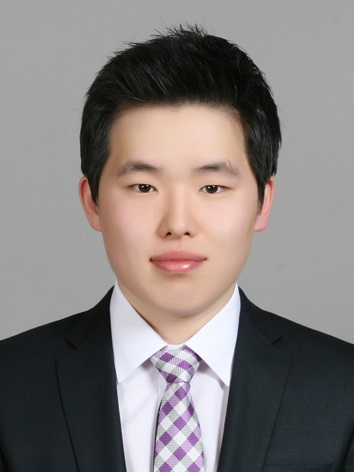}}]{Sungjun Seo} (Member, IEEE)
 received the B.S. and M.S. degrees in mechanical engineering from the Department of Mechanical Engineering, Kyungpook National University, Daegu, South Korea, in 2011 and 2013, respectively. He is currently pursuing the Ph.D. degree in mechanical engineering with the Department of Mechanical Engineering, New Mexico Institute of Mining and Technology, Socorro, NM, USA. He worked as a researcher with the Department of Vacuum Deposition Technology, LG Electronics Inc., Seoul, South Korea, from 2013 to 2017, and as a research staff member with the Department of Product Development, PCO Nhac Ltd., South Korea, from 2018 to 2022. His research interests include multi-agent systems, multi-agent area coverage, and wearable robots.\end{IEEEbiography}

\begin{IEEEbiography}[{\includegraphics[width=1in,height=1.25in,clip,keepaspectratio]{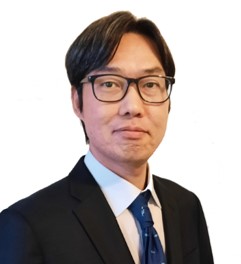}}]{Kooktae Lee} (Member, IEEE)
received the B.S. and M.S. degrees in Mechanical Engineering from Korea University, Seoul, South Korea, in 2006 and 2008, respectively, and the Ph.D. degree in Aerospace Engineering from Texas A\&M University, College Station, TX, USA, in 2015. From 2015 to 2016, he was a Postdoctoral Research Associate at Texas A\&M University, and from 2016 to 2017, he was a Postdoctoral Scholar in the Department of Mechanical and Aerospace Engineering at the University of California, San Diego. He joined the Department of Mechanical Engineering at the New Mexico Institute of Mining and Technology, Socorro, NM, USA, in 2017, where he is currently an Associate Professor. 

Dr.~Lee received the NSF CAREER Award in 2022 from the Division of Civil, Mechanical, and Manufacturing Innovation (CMMI), Dynamics, Control, and Systems Diagnostics (DCSD) program within the Directorate for Engineering (ENG). His research interests include robotics and control, multi-agent systems, distributed networked control, uncertainty quantification, asynchronous algorithms, and artificial intelligence.\end{IEEEbiography}

\end{document}